\documentclass[pr,twocolumn, superscriptaddress, showpacs, a4paper]{revtex4}

\usepackage{graphicx}

\begin{document}

\title{Spectral signatures of high-symmetry quantum dots and effects of symmetry breaking}

\author{K. F. Karlsson}
\email{freka@ifm.liu.se}
\affiliation{Ecole Polytechnique F\'ed\'erale de Lausanne (EPFL), Laboratory of Physics of Nanostructures, CH-1015 Lausanne, Switzerland}
\affiliation{Link\"oping University, Department of Physics, Chemistry, and Biology (IFM), \\Semiconductor Materials, SE-58183 Link\"oping, Sweden}

\author{D. Y. Oberli}
\affiliation{Ecole Polytechnique F\'ed\'erale de Lausanne (EPFL), Laboratory of Physics of Nanostructures, CH-1015 Lausanne, Switzerland}

\author{M. A. Dupertuis}
\affiliation{Ecole Polytechnique F\'ed\'erale de Lausanne (EPFL), Laboratory of Physics of Nanostructures, CH-1015 Lausanne, Switzerland}

\author{V. Troncale}
\affiliation{Ecole Polytechnique F\'ed\'erale de Lausanne (EPFL), Laboratory of Physics of Nanostructures, CH-1015 Lausanne, Switzerland}

\author{M. Byszewski}
\affiliation{Ecole Polytechnique F\'ed\'erale de Lausanne (EPFL), Laboratory of Physics of Nanostructures, CH-1015 Lausanne, Switzerland}

\author{E. Pelucchi}
\altaffiliation{Present address: Tyndall National Institute, University College Cork, Cork, Ireland.}
\affiliation{Ecole Polytechnique F\'ed\'erale de Lausanne (EPFL), Laboratory of Physics of Nanostructures, CH-1015 Lausanne, Switzerland}

\author{A. Rudra}
\affiliation{Ecole Polytechnique F\'ed\'erale de Lausanne (EPFL), Laboratory of Physics of Nanostructures, CH-1015 Lausanne, Switzerland}

\author{P.O. Holtz}
\affiliation{Link\"oping University, Department of Physics, Chemistry, and Biology (IFM), \\Semiconductor Materials, SE-58183 Link\"oping, Sweden}

\author{E. Kapon}
\affiliation{Ecole Polytechnique F\'ed\'erale de Lausanne (EPFL), Laboratory of Physics of Nanostructures, CH-1015 Lausanne, Switzerland}

\date{\today}

\begin{abstract}
A sequence of photoluminescence spectroscopy based methods are used to rigorously identify and study all the main spectral features (more than thirty emission lines) of site controlled InGaAs/AlGaAs quantum dots (QDs) grown along [111]B in inverted tetrahedral pyramids. The studied QDs reveal signatures of one confined electron level, one heavy-hole-like level and one light-hole-like level. The various heavy-light-hole hybrid exciton complexes formed in these QDs are studied by polarization resolved spectroscopy, excitation power dependence, crystal temperature dependence and temporal single photon correlation measurements. The presented approach, which only requires a minimal theoretical input, enables strict spectral identification of the fine structure patterns including weak and spectrally overlapping emission lines. Furthermore, it allows the involved electron-hole and hole-hole exchange interaction energies to be deduced from measurements. Intricate fine structure patterns are qualitatively understood by group theory and shown to be very sensitive to the exact symmetry of the QD. Emission patterns influenced by hole-hole exchange interactions are found to be particularly useful for identifying QDs with high $C_{3v}$ symmetry and for probing symmetry breaking.

\end{abstract}


\maketitle

\section{Introduction}
Epitaxial semiconductor quantum dots (QDs) that trap and confine electron and holes on a nanoscopic length scale have now been a subject of investigation for more than two decades. The initial driving force for the research was the application of QDs as the active medium in lasers, which potentially would reduce the lasing threshold current and improve the temperature stability of the devices \cite{JQE.22.1915}. More recently, the focus has shifted from the lasers towards the application of QDs as quantum light emitters in the area of optical quantum communication and information processing \cite{Nature.1.215}. In particular, the QDs have proven to be excellent emitters of single and time-correlated photons \cite{Nature.290.2282, PhysRevLett.87.183601}. Currently there is a significant effort to develop reliable QD-based emitters of entangled photons \cite{PhysRevLett.84.2513, PhysRevLett.96.130501, Nature.439.179}, to integrate QDs with photonics cavities and waveguides \cite{Nature.445.896} and to coherently manipulate the quantum states of the QDs \cite{Nature.456.218}. A detailed understanding about the optical properties of the QDs and their electronic structure is of utmost importance for this development.

The most widely studied QDs are so far the strained In(Ga)As dots formed in Stranski-Krastonov (SK) growth mode on (001) GaAs substrates. Numerous and detailed spectroscopy studies on these QDs addressing interband transitions have been published. A particular focus has been devoted to the exciton fine structure \cite{PhysRevB.65.195315, PhysRevB.67.161306}, since it can be detrimental for the generation of polarization entangled photons in the biexciton-exciton cascade decay \cite{PhysRevLett.96.130501, Nature.439.179}. Since the prediction of vanishing fine structure splitting for symmetric QDs grown in the [111] direction \cite{PhysRevB.81.161307, PhysRevLett.103.063601}, as the effect of three instead of two \cite{PhysRevB.71.045318} symmetry planes for such QDs, there has been a growing interest for alternative QD systems. Some alternative QD systems are based on nano-wire heterostructures \cite{NanoLett.5.1439}, droplet epitaxy \cite{APE.3.065203} and growth on non-planar substrates \cite{APL.71.1314}. In contrast to SK QDs, for which the formation mechanism relies on significant lattice mismatch between the substrate and the dot material, other approaches of fabrication can result in QDs with little or no strain. While the electronic properties of typical disc-like and strongly strained SK QDs can be fairly well understood by models that neglect valence band mixing and light-holes \cite{PhysRevB.65.195315}, this is not the case for some of the alternative QD systems for which such mixing is significant.

The purpose of this article is to present a complete analysis of the intrinsic spectral features of InGaAs QDs grown in inverted tetrahedral micropyramids etched into (111)B GaAs substrates \cite{APL.71.1314}. Unlike the typical SK QDs, these pyramidal QDs exhibit a significant degree of valence band mixing \cite{PhysRevB.81.161307, nl0706650}. The QDs of this study are of a high technological interest, since they ideally possess the high $C_{3v}$ symmetry required for generating polarization entangled photons \cite{PhysRevB.81.161307, NPhoton.7.527}. Moreover, they are site-controlled with extreme dot-to-dot uniformity in the spectral features \cite{APL.84.1943, APL.94.223121, Small.6.1268, APL.91.081106} and they are fully compatible with the present photonic crystal technology \cite{APL.92.263101}.  

For the analysis presented in this work, first the relevant carrier states are identified, followed by a careful identification of a total of eleven exciton complexes, exhibiting more than thirty emission lines. A combination of experimental methods is presented to rigorously identify the excitons complexes with minimal theoretical input and without any numerical modelling. We demonstrate that the approach allows a confident identification of all emission lines as well as strict prediction of missing emission lines hidden due to spectral overlap with other high intensity components. Thereafter, the observed dot-to-dot variation of the direct Coulomb interactions is qualitatively discussed, followed by a detailed analysis of the polarised fine structure patterns of all the identified exciton complexes. This analysis is based on symmetry arguments given by group theory alone and some general knowledge about the electronic structure of the QDs, and it enables extraction of all the relevant electron-hole and hole-hole exchange interaction energies from the experimental spectroscopic data. Finally, the effects of symmetry breaking are analyzed in terms of degeneracy lifting and relaxed optical selection rules. Most importantly, the spectral signatures of high symmetry QDs are identified. The methods presented in this work are general and they should be applicable to any QD system, and, in particular, the group theory approach for analysing the fine structure patterns is very effective when dealing with high symmetry QDs.

\section{Experimental details}
The experiments were performed on hexagonal arrays of QDs fabricated by low-pressure organometallic chemical vapor deposition (OMCVD) in tetrahedral recesses patterned with 5 $\mu$m pitch on a 2$^\circ$-off GaAs (111)B substrate \cite{APL.71.1314}. Thin QDs self-form due to growth anisotropy and capillarity effects \cite{PhysRevLett.81.2962, PhysRevB.57.R9416} from a nominally 0.5-nm-thick In$_{0.10}$Ga$_{0.90}$As layer at the center of the pyramids, sandwiched between Al$_{0.30}$Ga$_{0.70}$As barriers. The actual Al concentration in the barriers surrounding the QD becomes however reduced due to alloy segregation: A vertical quantum wire (VQWR) of nearly pure GaAs is formed along the vertical axis of the pyramid, intersecting the QD, and three vertical quantum wells with ~20\% Al concentration are connected to the QD and VQWR \cite{nl0706650, PhysRevB.82.165315}. Individual QDs were excited non-resonantly by a laser spot size of ~1 $\mu$m with the power in the range 25 - 750 nW and wavelength 532 nm. The sample was placed in a cold finger cryostat and kept at a constant temperature chosen in the range of 10 - 30 K. The single QD photoluminescence (PL) was split 50/50 and transferred to the entrance slits of two monochromators and recorded by either a charge-coupled device (CCD) or by avalanche photo diodes (APD), with the spectral resolution set to 110 $\mu$eV. The two APDs, operating in single-photon counting mode, terminated the two arms in a Hanbury Brown and Twiss setup and generated the start and stop signals for temporal single photon-correlation spectroscopy. The optical linear polarization of the luminescence was analyzed by a rotatable $\lambda/2$-plate and a fixed linear polarizer placed in the signal path in front of the entrance slit of the monochromator. The contrast in the polarization measurements was about 50:1.

The QDs were investigated both in the standard top-view geometry, with the luminescence extracted along the growth axis [$111$]B ($z$) enabling polarization analysis in the perpendicular $xy$-plane, and in a side-view geometry with the light extracted along [$1\bar{1}0$] ($x$) and the polarization analyzed in the $yz$-plane. For the measurements in side-view geometry, the samples were cleaved along [$11\bar{2}$] ($y$) and the QDs in partially cleaved pyramids were investigated \cite{APL.89.251113}. Note that the often ignored $z$-polarization, only accessible from the cleaved edge, is highly relevant for studies on QDs with valence band mixing. Before the measurements in the standard top-view geometry, the samples underwent a substrate-removal procedure called backetching which results in freestanding pyramids with significantly improved light extraction efficiency \cite{JPCM.11.5901}. In total, nearly 200 QDs were carefully analyzed in this study.

\section{Carrier states}

\begin{figure}[top] 
\includegraphics[width=\columnwidth]{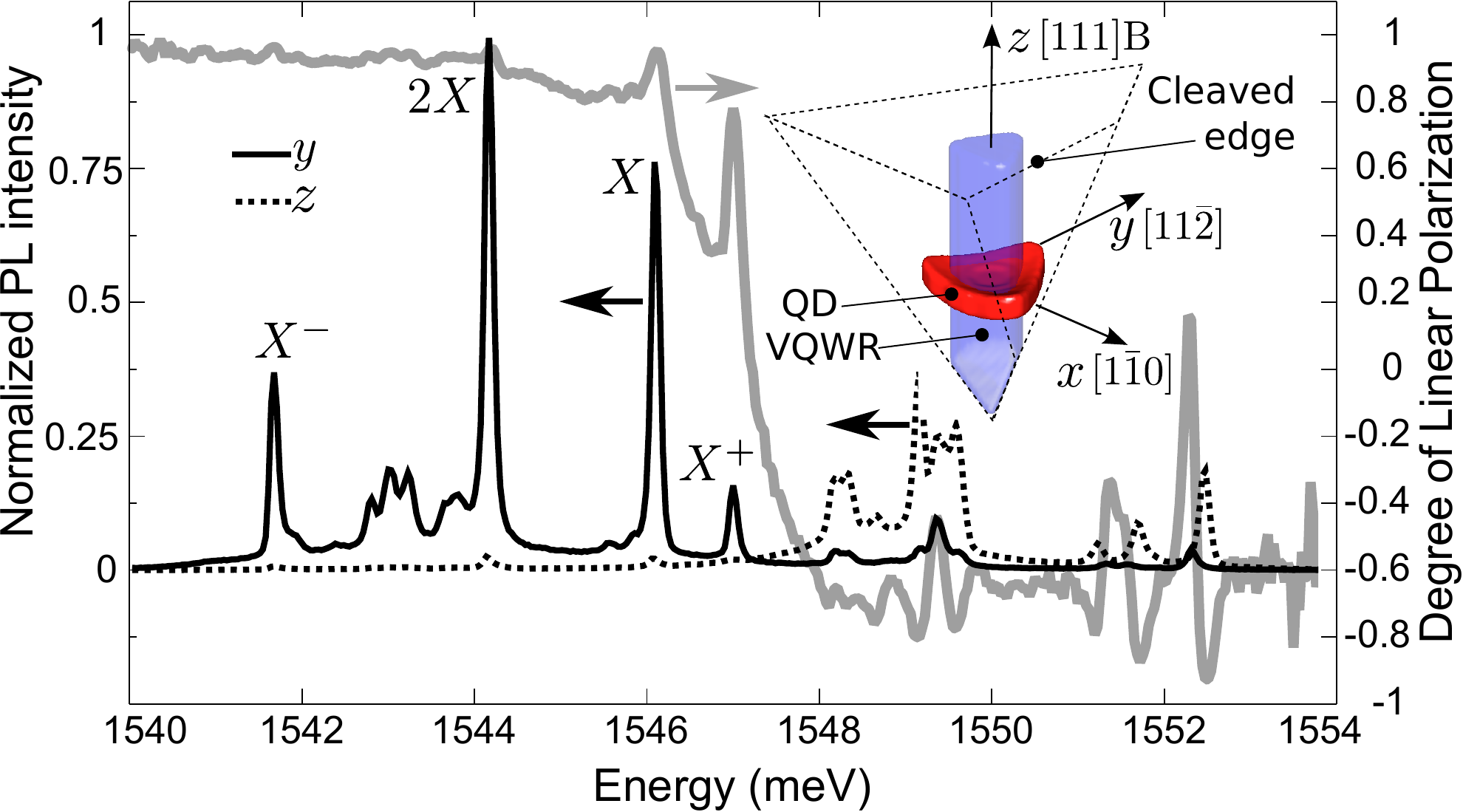}
\caption{\label{cleavedEdgePano} (Color online) Polarization resolved $\mu$PL spectra of a single QD acquired from the cleaved edge with 500 nW excitation power at $T$ = 27 K. The solid (dotted) black lines indicate PL intensity $I_y$ ($I_z$) linearly polarized along $y$ ($z$), the solid grey line indicates the degree of linear polarization $P = (I_y-I_z)/(I_y+I_z)$. The PL intensity is normalized with respect to $I_y$. The inset illustrates the QD and the VQWR in a partially cleaved inverted pyramid with the relevant crystallographic directions indicated.}
\end{figure}

The side-view interband PL emission of a single pyramidal QDs is rich of spectral features \cite{APL.89.251113, PhysRevB.81.161307}. Identifications of the various emission lines in terms of exciton complexes will be presented in the following sections. In this section the focus is on the fact that the emission lines can be divided into two groups distinguished by their linear polarization: Group 1 corresponds to polarization vectors in the $xy$-plane and group 2 is mainly polarized in the perpendicular $z$-direction. This is revealed in the typical side-view spectrum shown in Fig. 1, where also the degree of linear polarization ($P$) is plotted. Note the abrupt transition in the average degree of polarization from $P\approx +1$ to $P\approx -0.6$ when the photon energy passes from group 1 to group 2.

The valence band eigenstates at the Brillouin zone center for typical zincblende quantum wells (QWs) can have either of two characters of the Bloch-periodic part of the wave function, refered to as heavy- and light-holes. In a quantum well the optical interband transitions between an electron in the conduction band and a heavy-hole (hh) exhibit $P = +1$, while the corresponding value is $P = -0.6$ for a light-hole (lh). It is therefore natural to refer to the QD emission lines belonging to groups 1 and 2 as well as the corresponding hole states as heavy-hole-\emph{like} and light-hole-\emph{like}, respectively, although the single-particle eigenstates in a QD do not necessarily have pure hh or lh characters. In fact, $k \cdot p$ modelling of the pyramidal QDs yields about 10\% lh in the hh-like states and about 10\% hh in the lh-like states \cite{PhysRevB.81.161307}. Similar to QWs are the hh-like transitions here being observed with lowest energy while the lh-like transitions appear at higher energies. This indicates that the hole ground level is hh-like and that the excited energy level is lh-like.

The relative PL intensity of the lh-like emission increases with the crystal temperature $T$, as shown in Fig. 2(a). Above $T$ = 25 K, the relative intensity of lh-like emission corresponds to an activation energy $E_A$ = $\sim$ 7 meV, which is in fair agreement with the observed energy separation of $\sim$ 9  meV between the dominating pairs of emission lines with lh-like and hh-like polarization [see Fig. 2(b)]. The discrepancy of nearly 2  meV can partly be understood as different Coulomb interaction energies between the electron and a hole either in the hh-like or in the lh-like state, as will be discussed in Section V. It can therefore be concluded that the energy separation between the hh-like and lh-like states must be of a similar magnitude ($\sim$ 7 to 9 meV), and that the energetically higher lh-like hole states essentially are thermally populated for $T \geq 25$ K. For $T < 25$ K, on the other hand, it is clear from the Arrhenius plot in Fig. 2(b) that the intensity of the lh-like emission is significantly stronger than expected from purely thermal excitation. This indicates a reduced relaxation efficiency of the holes from the lh-like state down to the hh-like. This effect may be the result of the acoustic phonon relaxation bottleneck which has been predicted in zero-dimensional system by Bockelmann and Bastard \cite{PhysRevB.42.8947}, and observed subsequently in single GaAs/AlGaAs QDs by Brunner \emph{et al.} \cite{PhysRevLett.69.3216}.

There are no spectral evidences of any additional hole level confined to the QD, besides the discussed hh- and lh-like levels. Moreover, there are no indications of any excited electron levels in the spectra. Consequently, all the details of the pyramidal QD spectra will in the following be analyzed on the basis of one single electron level, one hh-like level and one lh-like level. Each level is assumed to accommodate two degenerate single particle states due to Kramer's degeneracy.

\begin{figure}[top]  
\includegraphics[width=\columnwidth]{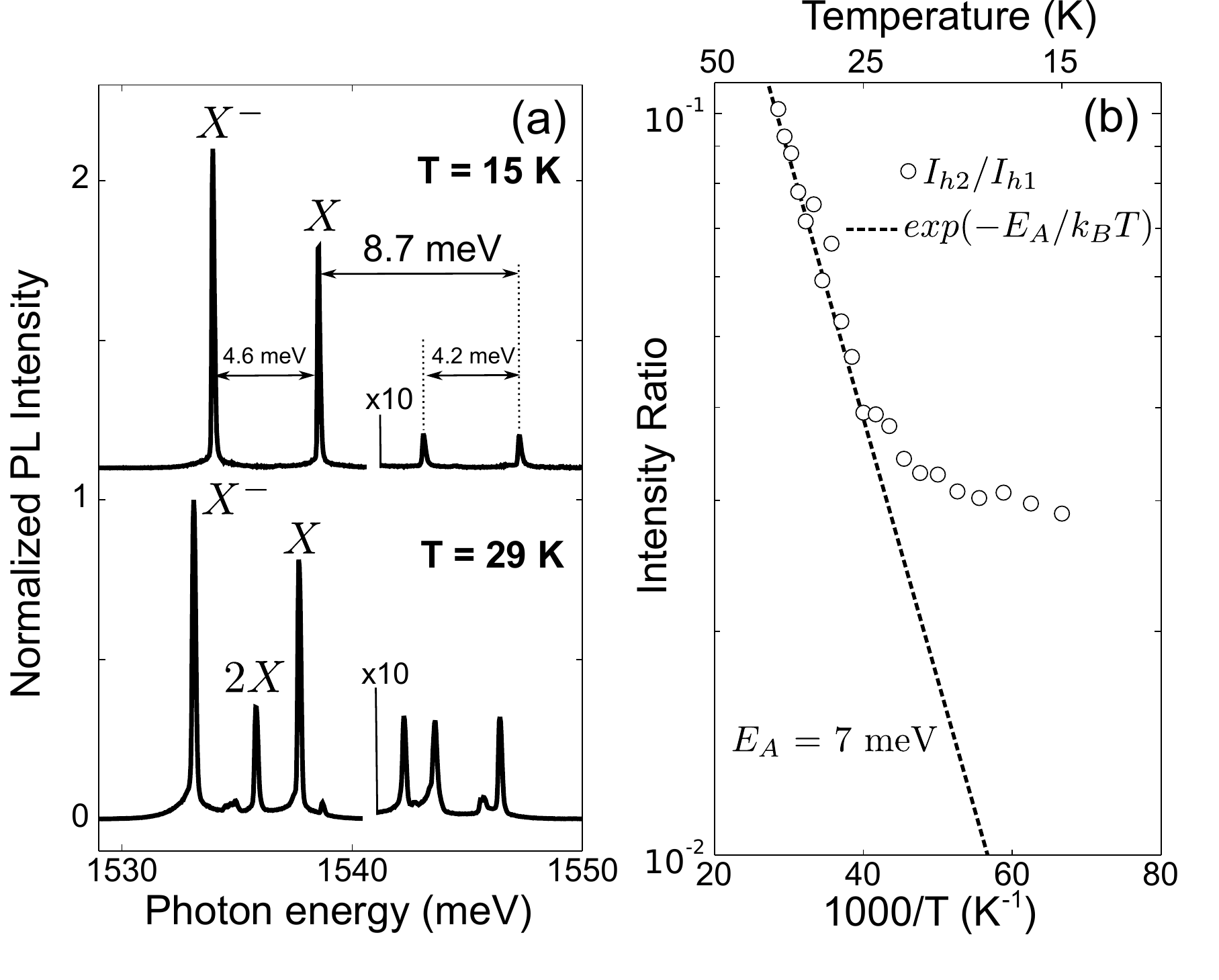}
\caption{(a) $\mu$PL spectra of a single QD acquired in top view at two different temperatures. The lh-like emission is enhanced by a factor 10. (b) The ratio between the total integrated intensity of the lh-like emission lines ($I_{h2}$) and the hh-like emission lines ($I_{h1}$) is shown with circles and the Boltzmann factor corresponding to activation energy $E_A$ = 7 meV is shown as a dashed line.}
\label{activation}
\end{figure}

\section{Exciton complexes}

Electrons (e) and holes (h) confined in a QD under the influence of mutual Coulomb interactions form a large variety exciton complexes. Exciton complexes have been identified and investigated for various QD systems, and for the most well-studied complexes,  all the electrons and holes occupy the corresponding ground state levels, such as the single exciton X (1e1h), the biexciton 2X (2e2h), the negative trion X$^-$ (2e1h) and the positive trion X$^+$ (1e2h). For the pyramidal QD system, these complexes dominate the spectra measured in the standard top-view geometry, corresponding to the solid black line in Fig. 1.  Their spectral identification (see the labels in Figs. 1 and 2) have previously been firmly justified by temporal photon correlation measurements \cite{PhysRevB.73.205321}.

The charge state of the complexes in a pyramidal QD can conveniently be optically controlled by the excitation power, according to an established photo-depletion model \cite{PhysRevLett.84.5648}. For the majority of the pyramidal QDs, but not all, the dominating charge state also can be controlled by the crystal temperature \cite{APL.78.2952, JAP.101.081703}.  The coexistence of exciton complexes with different charge states in the same spectrum is an inherent property of a system populated by random and separate processes of electron and hole capture \cite{PhysRevB.55.9740, PhysRevB.73.205321}.

\subsection{Notation}
\begin{figure}[top]   
\includegraphics[width= 8.5 cm]{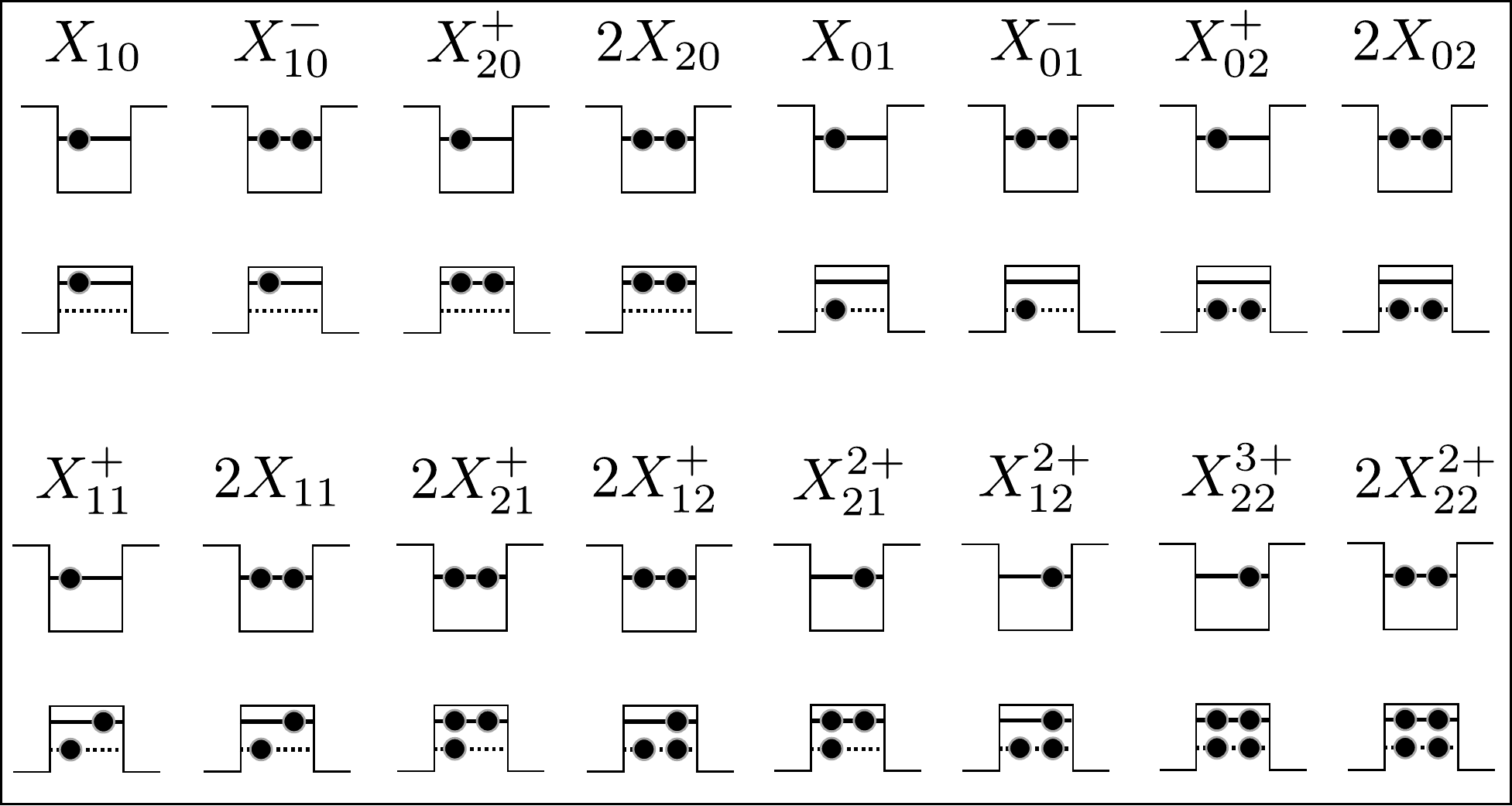}
\caption{ Schematic diagrams of the electronic configurations of exciton complexes for shallow-potential pyramidal QDs. The lh-like hole level is represented by a horizontal dotted line.}
\label{exciton_schemes}
\end{figure}

The above mentioned electron and hole levels in the pyramidal QDs enable the existence of 16 different exciton complexes. In order to be able to address a specific complex unambiguously, we extend the standard labeling with two subscript indices, where the first (second) index denotes the number of holes occupying the first (second) hole level \cite{PhysRevB.81.161307}. Schematic diagrams of all the possible exciton complexes together with their labels are shown in Fig. ~\ref{exciton_schemes}.

The optical decay of an exciton complex will produce different spectral outcomes depending on whether the recombining hole is hh-like or lh-like. For a complex involving holes in both levels, e.g.  $X^+_{11}$, two different optical transitions exists depending on if it is the hh-like or lh-like hole that recombines. The two possibilities will be distinguished by a bar above the index in the label representing the energy level of the recombining hole. Thus, for the given example, $X^+_{\bar{1}1}$ indicates the recombination of the hole in the hh-like level and $X^+_{1\bar{1}}$ that of the lh-like level.

\subsection{Experimental identification}

\begin{figure}[top] 
\includegraphics[width= 8 cm]{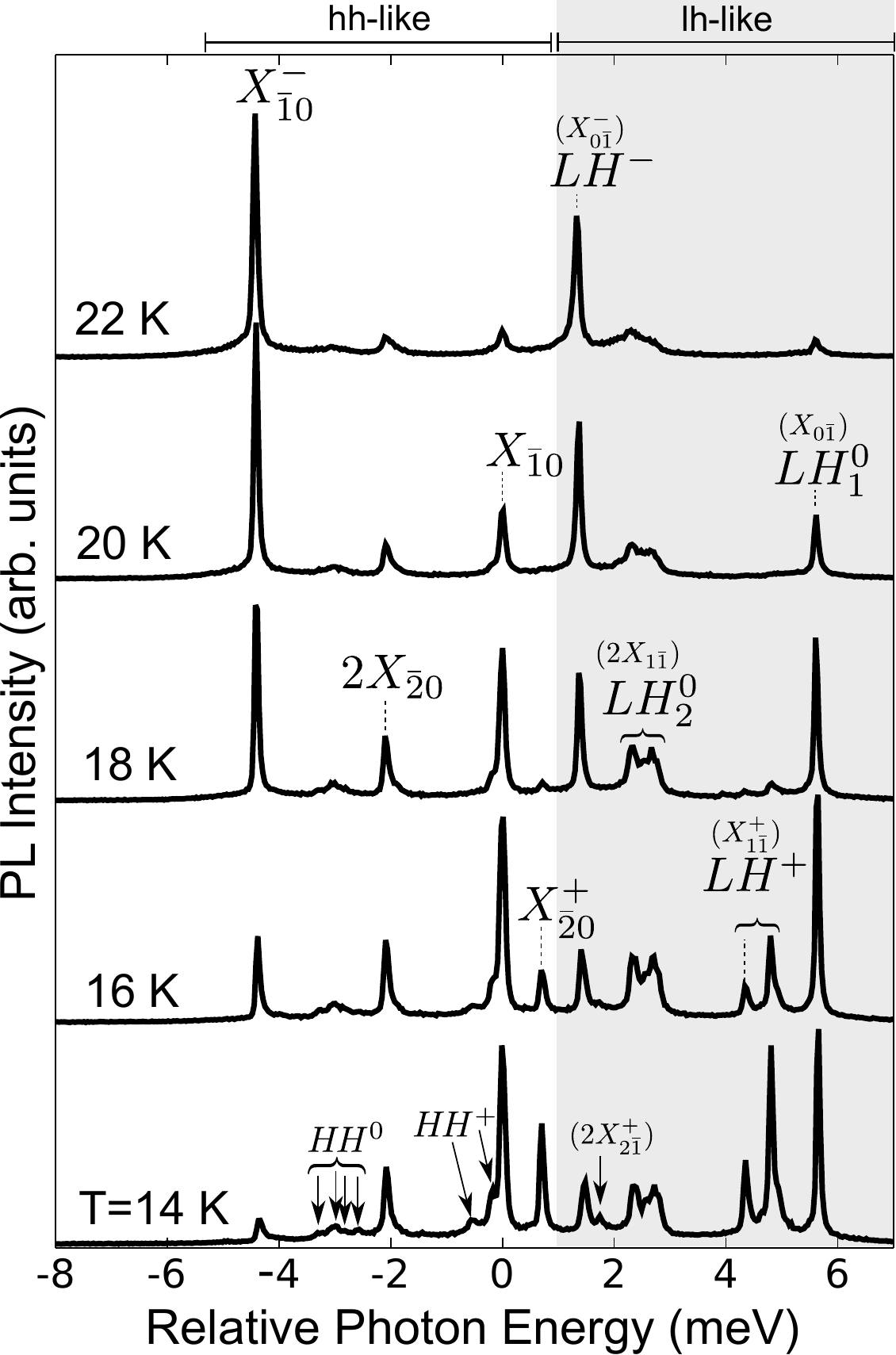}
\caption{ $\mu$PL spectra of a single QD acquired from the cleaved edge at different crystal temperatures. The detected polarization was set to 60$^\circ$ with respect to $y$ so that both hh-like and lh-like (shaded area) transitions could be detected with comparable intensities. A laser power of 35 nW was used for the measurements. }
\label{temp_spc}
\end{figure}

\begin{figure}[top]  
\includegraphics[width= 8 cm]{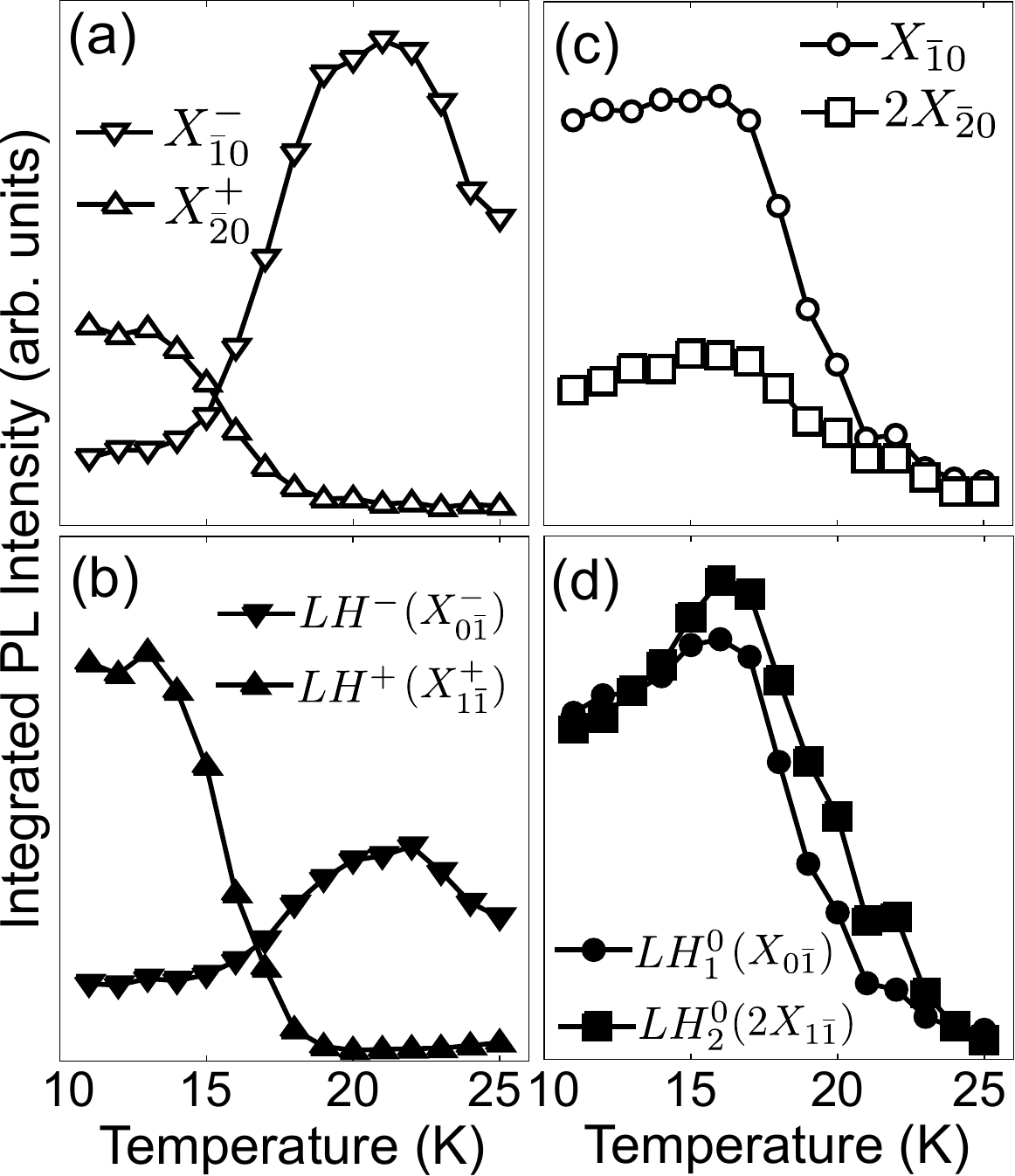}
\caption{ Temperature dependence of the PL intensity corresponding to (a) and (b) the positively and negatively charged exciton complexes and (c) and (d) neutral exciton complexes.}
\label{temp_curve}
\end{figure}

The dominating lh-like emission lines can be identified by correlating their intensities with the well known hh-like emission lines ($X_{\bar{1}0}$, $2X_{\bar{2}0}$, $X^-_{\bar{1}0}$ and $X^+_{\bar{2}0}$), as the charge population of the dot is tuned by a parameter like excitation power or crystal temperature. Fig. ~\ref{temp_spc} shows spectra at different temperatures for a typical QD, measured from the cleaved edge with the polarization analyzer in an intermediate position transmitting both hh-like and the lh-like emission lines with comparable intensities \cite{JAP.101.081703}.  It is clear that the left part of the spectrum for this QD, corresponding to hh-like transitions, undergoes dramatic changes with the temperature: As the temperature is increased the average charge state of the dot successively changes from mainly positive, indicated by the strong intensity of $X^+_{\bar{2}0}$, to largely negative with intensity of $X^-_{\bar{1}0}$ becoming dominant. The full temperature evolutions of $X^-_{\bar{1}0}$ and $X^+_{\bar{2}0}$ are plotted in Fig.  ~\ref{temp_curve} (a). Emission lines with temperature dependencies very similar to $X^-_{\bar{1}0}$ or $X^+_{\bar{2}0}$ can also be identified in the right lh-like part of the spectrum, as shown by the single line denoted LH$^-$ and the group of lines denoted LH$^+$ in Figs. 4 and 5 (b). This suggests that LH$^-$ originates from a negatively charged complex, while the group LH$^+$ is related to one or several positively charged complexes. Note that the unique temperature dependence of the hh-like emission lines $X_{\bar{1}0}$ and $2X_{\bar{2}0}$, originating from neutral complexes, is also reproduced for some of the lh-like emission lines, denoted LH$^0_1$ and LH$^0_2$ in Figs ~\ref{temp_spc} and ~\ref{temp_curve}.

According to Fig. 3, there is merely one way to form a negative trion with lh-like emission lines, $X^-_{0\bar{1}}$, for the investigated shallow-potential pyramidal QDs. Hence, the single emission line LH$^-$ is identified as $X^-_{0\bar{1}}$.

The intensity of LH$^+$ is the only part of the lh-like emission that correlates well with the intensity of the positive trion. However, there are two ways in which trions with lh-like emission can be generated, either by $X^+_{0\bar{2}}$ or by $X^+_{1\bar{1}}$ (see Fig. ~\ref{exciton_schemes}). The former case requires two excited holes, while only one hole is excited for $X^+_{1\bar{1}}$. It can therefore be argued that $X^+_{1\bar{1}}$, and not $X^+_{0\bar{2}}$, should dominate the spectra, since the temperatures used in these experiments correspond to a thermal energy of 1-2 meV that is much smaller than than the energy spacing between the hole levels ($\sim$ 7 to 9 meV). Thus, it is significantly less likely to find a complex populated with two excited holes, as compared to a complex populated with one excited hole. This does not exclude that some weak emission from $X^+_{0\bar{2}}$ also could be present in the spectra. The tentative assignement of LH$^+$ to the complex $X^+_{11}$ will be verified and confirmed later in this section.

The two lh-like emission lines LH$^0_1$ and LH$^0_2$ which correlates with neutral complexes can be related to either the single exciton $X_{0\bar{1}}$ or the biexcitons $2X_{1\bar{1}}$ and $2X_{0\bar{2}}$. Again, the temperature argument given above speaks in the favor of $2X_{1\bar{1}}$ as the most dominating biexciton related emission, instead of $2X_{0\bar{2}}$ with two excited holes. Therefore, LH$^0_1$ and LH$^0_2$ are tentatively attributed to the single exciton $X_{0\bar{1}}$ and the biexciton $2X_{1\bar{1}}$, respectively. This attribution will be verified in the following.

It is interesting to note that both LH$^+$ and LH$^0_2$ consist of a set of at least three resolved emission lines, all within a narrow spectral range of $\sim$ 0.4 meV. This implies that the maximal interaction energy involved to induce this fine structure are of the same order $\sim$ 0.4 meV. As this energy is well below the thermal energy, the spectral fine structure components of the corresponding complexes should be nearly independent of the temperature. This is confirmed by experiments; while the temperature dependent charging dramatically redistributes the PL intensities among different exciton complexes, the relative intensities of multiple emission lines belonging to the same complex is essentially remaining constant. Thus, this approach can be consistently used to determine exactly which of the emission lines are related to a certain complex, and which of them are not.

\begin{figure}[top]  
\includegraphics[width= 8.5 cm]{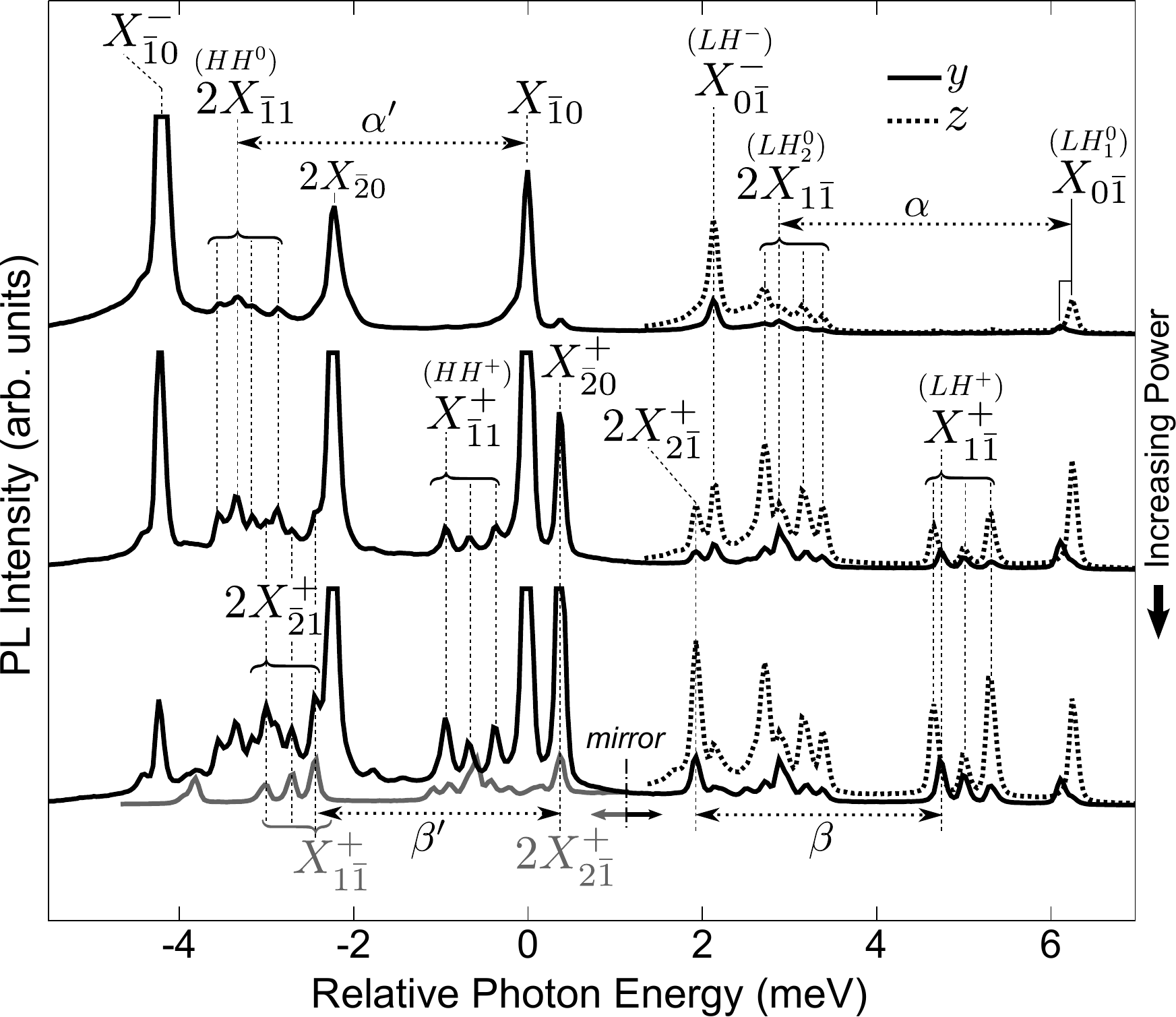}
\caption{Polarization resolved $\mu$PL spectra of a single QD acquired  at $T$ = 27 K from the cleaved edge with successively increasing excitation powers up to 500 nW. The solid (dotted) lines indicate PL intensity linearly polarized along $y$ ($z$). Dominating emission lines saturate the detector.}
\label{power_dep}
\end{figure}

The identification of the lh-like emission lines $2X_{1\bar{1}}$ and $X^+_{1\bar{1}}$ should be made in comparison with the corresponding hh-like transitions $2X_{\bar{1}1}$ and $X^+_{\bar{1}1}$. Potential candidates for such hh-like emission lines are indeed seen as weak features marked by arrows at the bottom of Fig. ~\ref{temp_spc}. In particular, the set denoted HH$^0$ correlates in intensity with the neutral complexes while the set HH$^+$ follows the temperature dependence of the positive trion. These weak features are more clearly seen in the power dependent spectra of a different QD presented in Fig. ~\ref{power_dep}, where the dominating peaks cause saturation of the detector. Analogous to the temperature dependence, the dot progressively changes its charge state from being negative at low powers to be become mainly positive at high powers.

\begin{figure}[Bottom] 
\includegraphics[width= 8 cm]{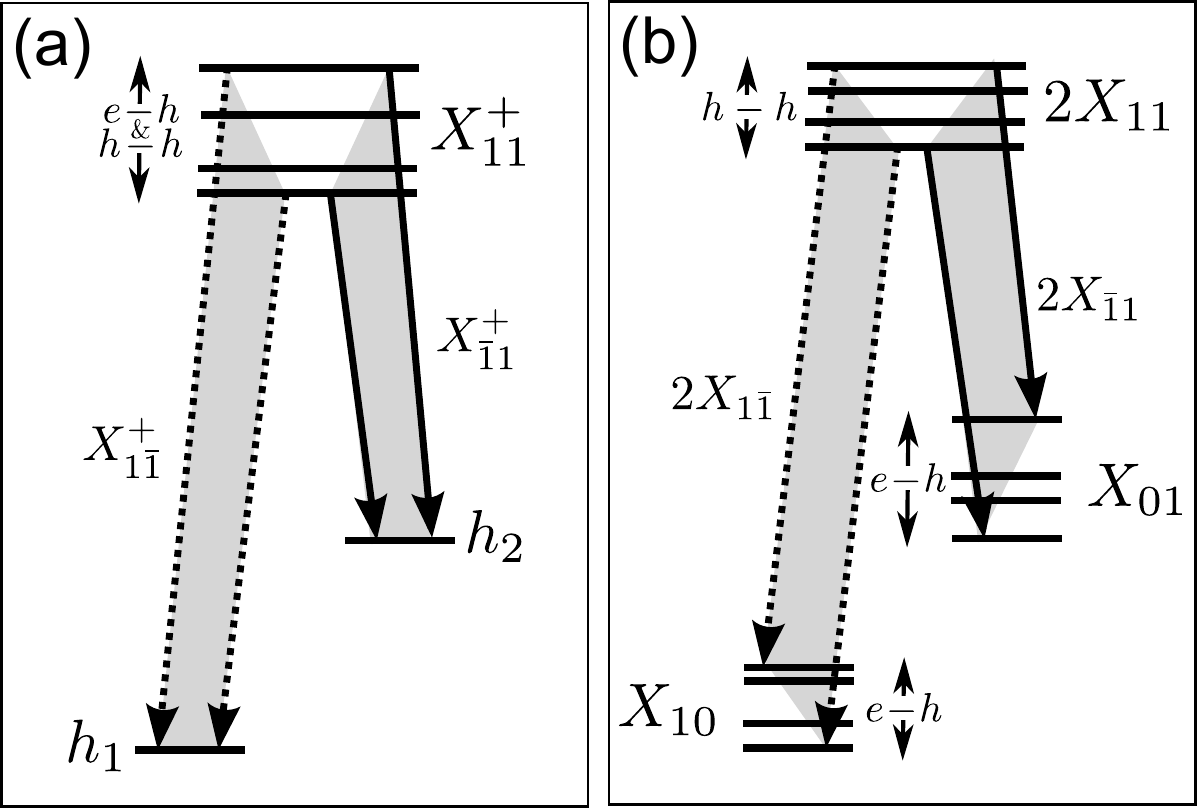}
\caption{ Decay diagrams of (a) the positive trion $X^+_{1\bar{1}}$ and (b) the biexciton $2X_{1\bar{1}}$. Fine structure splittings caused by e-h and h-h exchange Coulomb interactions will be described in detail in section VI.}
\label{dec_2X11_Xp11}
\end{figure}

As the power increases, three lh-like emission lines previously attributed to $X^+_{1\bar{1}}$ appear in each polarization resolved spectrum. Simultaneously, three well resolved hh-like lines appear on the low-energy side of the single exciton $X_{\bar{1}0}$. These three lines exhibit identical energy spacings as the three $y$-polarized emission lines of $X^+_{1\bar{1}}$. Identical energy spacings for $X^+_{1\bar{1}}$ and $X^+_{\bar{1}1}$ are indeed expected, since the final state of both transitions is a sole hole, occupying a degenerate ground state (in absence of external magnetic field). Thus, all splittings occur in the initial states of $X^+_{11}$, which is identical for both transitions as illustrated in Fig.  ~\ref{dec_2X11_Xp11} (a). Consequently, the set of lines appearing on the low-energy side of $X_{\bar{1}0}$ are attributed to $X^+_{\bar{1}1}$ (see Fig ~\ref{power_dep}). Unlike the QD presented in Fig. ~\ref{power_dep}, the $X^+_{\bar{1}1}$ transitions are less clearly resolved for most of the studied dots due to partial spectral overlap with the strong emission line $X_{\bar{1}0}$.

For the QD shown in Fig ~\ref{power_dep}, the biexciton complex $2X_{1\bar{1}}$ corresponds to four well resolved emission lines. A corresponding set of lines that can be attributed to $2X_{\bar{1}1}$ are observed at low powers on the low-energy side of $2X_{\bar{2}0}$. In this case, the details of the spectral patterns of $2X_{1\bar{1}}$ and $2X_{\bar{1}1}$ are expected to be different, as a consequence of unequal splittings of the final states characterizing the hh- and lh-like transitions [see Fig. ~\ref{dec_2X11_Xp11} (b)]. 

\begin{figure}[top] 
\includegraphics[width= 8 cm]{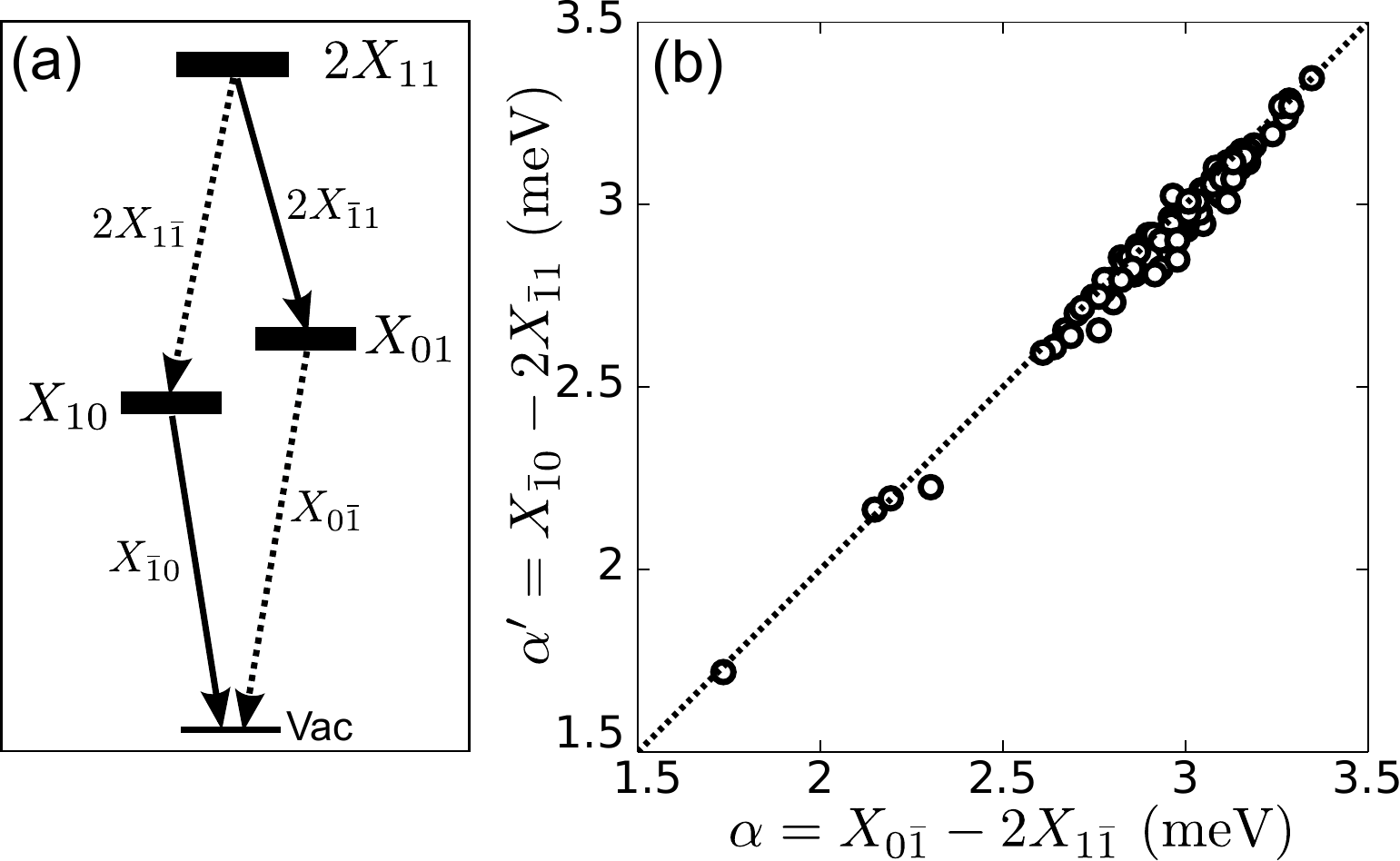}
\caption{(a) Diagram of the cascade decay of $2X_{11}$. Thick horizontal lines represent a group of energy levels while the thin horizontal line represents a single energy level. (b) Plot of the energy separations $\alpha'$ and $\alpha$ (defined in Fig. \ref{power_dep}) extracted from the experimental spectra of 168 QDs.}
\label{ecorr_alpha}
\end{figure}

A very strong support for the proposed attributions to $2X_{1\bar{1}}$ and $2X_{\bar{1}1}$ as well as $X_{0\bar{1}}$ can be derived from the full cascade of radiative decay from $2X_{11}$ to an empty dot, as shown in Fig. ~\ref{ecorr_alpha} (a), relying on the fact that the initial states $2X_{11}$ and the final state (empty QD) are identical for both decay paths \cite{PhysRevB.72.195332}. Neglecting any fine structure splittings of the intermediate single excitons, the global energy separation $\alpha'$ = ${X_{\bar{1}0}}$ - ${2X_{\bar{1}1}}$ must be approximately equal to the global energy separation $\alpha$ = ${X_{0\bar{1}}}$ - ${2X_{1\bar{1}}}$. Furthermore, with precise knowledge about the excitonic fine structure, exact energy relations applies to some specific transitions (see section VII). However, at this stage of the analysis the identification of the spectral intervals between the appropriate transitions $\alpha$ and $\alpha'$ given in Fig. ~\ref{power_dep} indeed yield $\alpha  \approx \alpha' \approx $ 3.35 meV for this particular QD. This energy relation also holds for all studied dots for which the relevant emission lines could be unambiguously identified (i.e. no spectral overlap with other transitions), and measured values of $\alpha$ and $\alpha'$ are reported in Fig. ~\ref{ecorr_alpha} (b) for 168 QDs.

\begin{figure}[Bottom] 
\includegraphics[width= 8 cm]{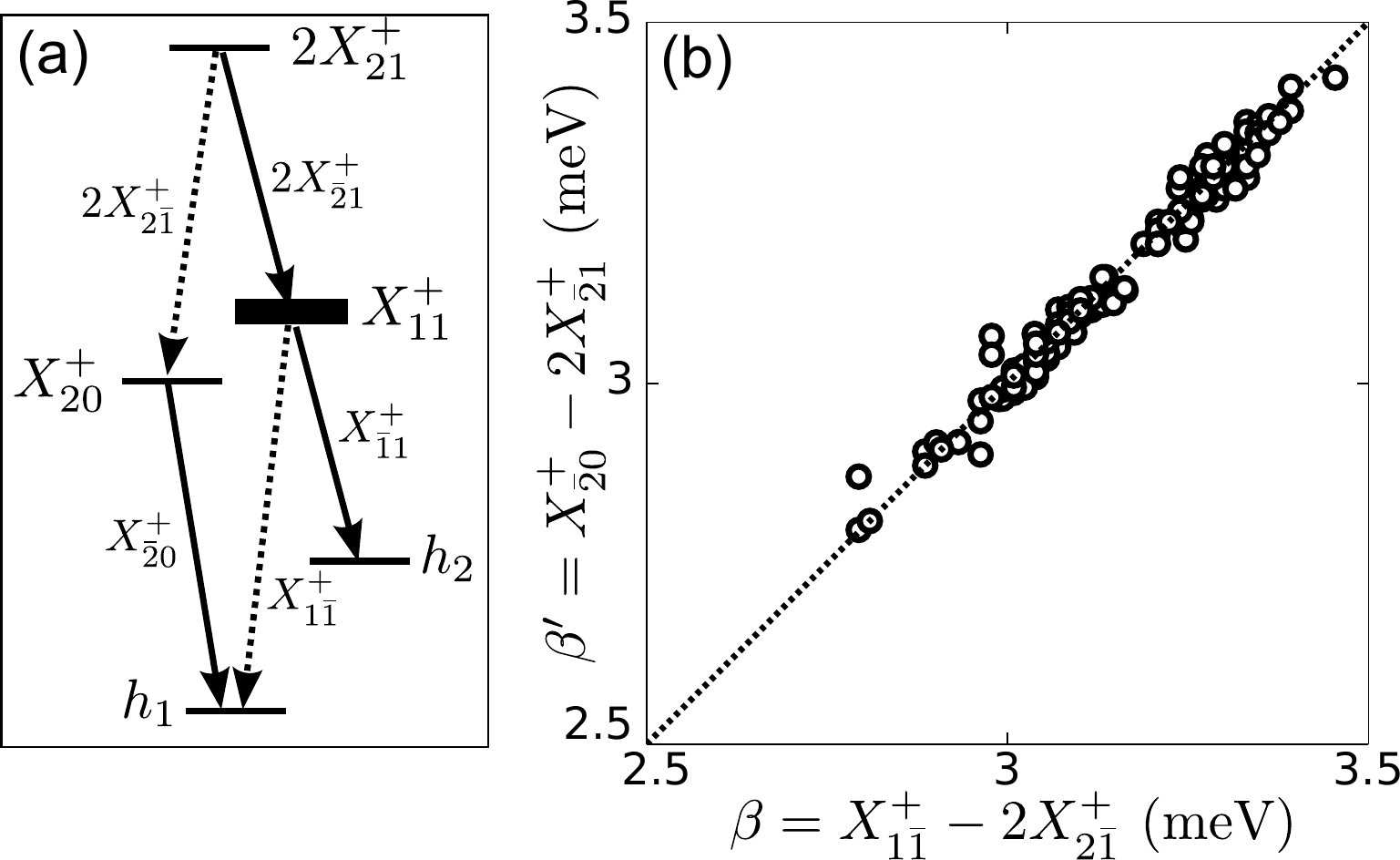}
\caption{\label{Fig. 8} Diagram of the cascade decay of $2X^+_{21}$. Thin horizontal lines represent a single energy levels while the thick horizontal line represents a group of energy levels. (b) Plot of the energy separations $\beta'$ and $\beta$ (defined in Fig. \ref{power_dep}) extracted from the experimental spectra of 189 QDs.}
\label{ecorr_beta}
\end{figure}

As the power is increased (see Fig. ~\ref{power_dep}), a new set of hh-like emission lines appear on the low-energy side of $2X_{\bar{2}0}$, superimposed on $2X_{\bar{1}1}$. Concurrently, a new single lh-like emission line appears on the low-energy side of $X^-_{0\bar{1}}$. These new transitions can confidently be attributed to a positively charged biexciton, namely $2X^+_{2\bar{1}}$ and $2X^+_{\bar{2}1}$, as explained in the following. The full decay scheme of $2X^+_{21}$ is shown in Fig. ~\ref{ecorr_beta} (a), and in analogy to the discussion above about the biexciton $2X_{11}$, there are two paths in the cascade decay with identical initial and final states. However, in this case only one of the two intermediate states is split, and exact energy relations can therefore be derived without any further detailed knowledge about the fine structure of the intermediate state. With $\beta = {X^+_{1\bar{1}}} - {2X^+_{2\bar{1}}}$ and $\beta' = {X^+_{\bar{2}0}} - {2X^+_{\bar{2}1}}$, one finds from Fig. ~\ref{ecorr_beta} (a) the strict relation $\beta' = \beta$. This implies that the spectral pattern of $2X^+_{\bar{2}1}$ essentially is the energetically mirrored pattern of $X^+_{1\bar{1}}$. The spectral position of the mirror is defined in Fig. ~\ref{power_dep} at the half of the distance between the spectral lines corresponding to  $X^+_{\bar{2}0}$ and $2X^+_{2\bar{1}}$. The strict relation $\beta' = \beta$ is verified for the QD in Fig. ~\ref{power_dep}, and is clearly illustrated by the mirrored $y$-polarized lh-like spectrum that is displayed as a grey line at the bottom. It is also verified for all the other investigated QDs that allowed unambiguous identifications as summarized in Fig. ~\ref{ecorr_beta} (b). Although the spectrum of $X^+_{1\bar{1}}$ for the QD shown in Fig. ~\ref{power_dep} consists of four components, only three of them are present in the spectrum of $2X^+_{\bar{2}1}$. The fourth component at lowest energy is unresolved because of its weak intensity and spectral overlap with the next stronger component at higher energy.

Note that $2X^+_{2\bar{1}}$ is very different from the other observed lh-like transitions, in the sense that this complex is in its ground state. $2X^+_{21}$ can be created at high excitation powers by state filling, while all the other lh-like transitions discussed here require the excitation of holes.

\begin{figure}[top] 
\includegraphics[width= 8 cm]{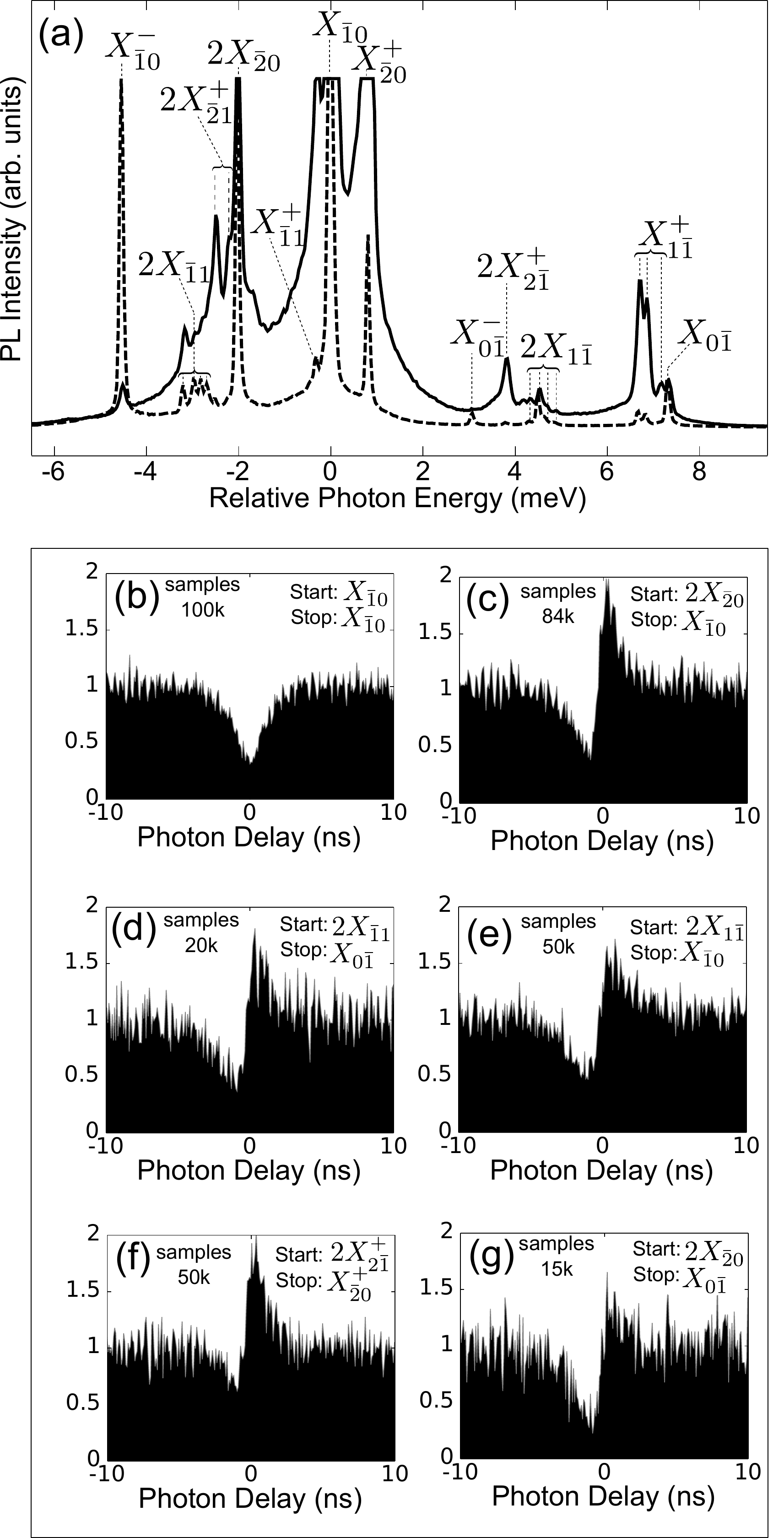}
\caption{ (a) $\mu$PL spectra of a single back-etched QD acquired in the standard top-view geometry with 75 nW (solid line) and 92 nW (dashed line) excitation powers at $T$ = 28 K. Dominating emission lines saturate the detector. (b) to (g) Measured second order photon correlation functions. The start and stop signals as well as the total number of coincidences are indicated in each histogram.}
\label{tcorr}
\end{figure}

As a final confirmation of peak identification conducted in this section, second order photon correlation spectroscopy will be employed for the cascade decays illustrated in Figs. ~\ref{ecorr_alpha} (a) and ~\ref{ecorr_beta} (a). Spectra of the QD used for the correlation measurements are shown in Fig. ~\ref{tcorr} (a). A conventional symmetric anti-bunching dip is observed in the measured auto-correlation function at zero time delay, obtained by  $X_{\bar{1}0}$ photons as both the start and the stop signals [see Fig. ~\ref{tcorr} (b)], simply reflecting the single photon emission property of the QD. For a biexciton-exciton cascade decay, photon bunching is revealed in the cross-correlation function for which a photon from the biexciton generates the start signal and the subsequently emitted exciton photon triggers the stop. Such a typical cross-correlation function exhibiting photon bunching is shown in Fig. ~\ref{tcorr} (c) for the conventional biexciton $2X_{\bar{2}0}$ and the single exciton $X_{\bar{1}0}$. Similar cross-correlation functions are expected for any biexciton-exciton cascade decay, and the photon bunching revealed in Fig. ~\ref{tcorr} (d) for $2X_{\bar{1}1}$ and $X_{0\bar{1}}$ as well as in Fig. ~\ref{tcorr} (e) for $2X_{1\bar{1}}$ and $X_{\bar{1}0}$ confirms the previous identification of these transitions. Moreover, the cascade decay of $2X^+_{21}$ is confirmed in Fig. ~\ref{tcorr} (f), for the decay path $2X^+_{2\bar{1}} \rightarrow X^+_{\bar{2}0}$. The intensity of $2X^+_{\bar{2}1}$ is too weak to allow cross-correlation measurements with any of the two alternative decay paths $2X^+_{\bar{2}1} \rightarrow X^+_{\bar{1}1}$ and $2X^+_{\bar{2}1} \rightarrow X^+_{1\bar{1}}$. Finally, Fig. ~\ref{tcorr} (g) shows that bunching is also observed for $2X_{\bar{2}0}$ and $X_{0\bar{1}}$, indicating that although the recombination of $2X_{20}$ prepares the QD with a single exciton in the ground state $X_{10}$, this exciton may subsequently recombine from its excited state $X_{01}$. This result is consistent with a thermal population of the excited lh-like level, which was previously demonstrated to occur at $T \geq 25$ K.

\begin{figure}[top] 
\includegraphics[width = 8.5 cm]{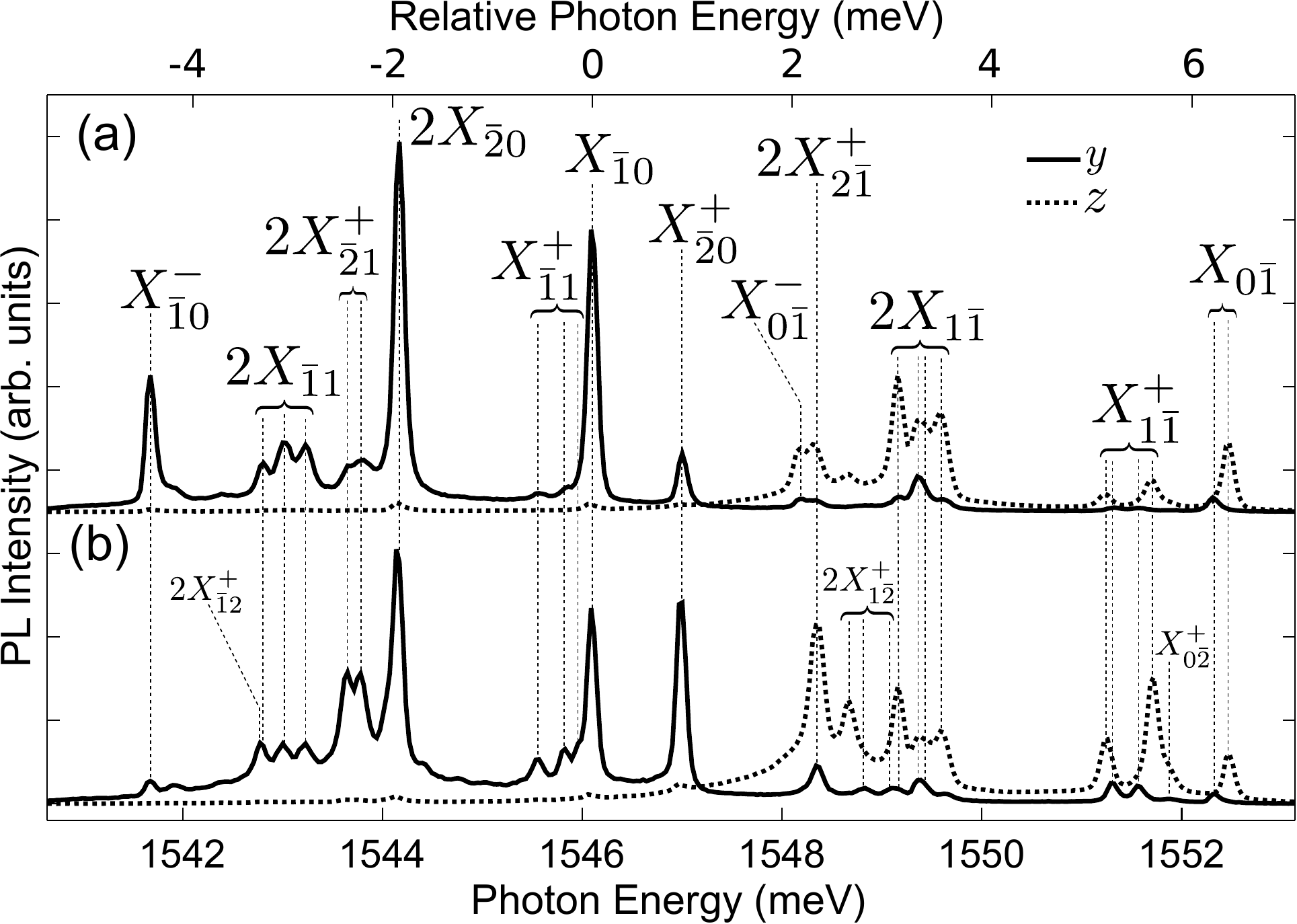}
\caption{$\mu$PL spectra of a single QD acquired from the cleaved edge with (a) 500 nW (b) 650 nW excitation powers at $T$ = 27 K. Solid (dotted) black curves represent $y$-polarized ($z$-polarized) emission. Experimentally rigorously identified excitonic transitions are indicated by big labels, while the small label in the bottom spectrum are tentative attributions consistent with the analysis in section VII.}
\label{pano_CE_symQD}
\end{figure}

In conclusion of this part, the spectral features pertaining to the very same exciton complex, as well as the charge state of the complex, were identified by unchanged relative intensities of the emission lines upon controlled charging. Thereafter, strict relations among certain spectral lines and their energies were applied in order to firmly attribute a set of emission lines to a specific exciton complex. For transitions with strong enough intensity, the identification was further confirmed with single photon temporal correlation spectroscopy. Figure ~\ref{pano_CE_symQD} (a) summarizes all the excitonic emission lines that have so far been rigorously identified for the investigated shallow-potential pyramidal QDs.

In figure ~\ref{pano_CE_symQD} (b) we report additional weaker emission lines appearing in between $2X^+_{2\bar{1}}$ and $2X_{1\bar{1}}$ at high excitation powers. There is also a weak shoulder on the high energy side of $X^+_{1\bar{1}}$. The attribution of these features is not as certain as those given in Fig. ~\ref{pano_CE_symQD} (a), but we will show in section VII that an identification of these features as $2X^+_{1\bar{2}}$ and $X^+_{0\bar{2}}$ is fully consistent with the experimental data and with the expected fine structure splitting patterns of these complexes. Furthermore, this identification implies that $2X^+_{\bar{1}2}$ spectrally overlaps with $2X^+_{\bar{1}1}$, which explains the change of the relative intensities of the components of this complex at high excitation powers. 

\section{Coulomb energies}

\begin{figure}[top] 
\includegraphics[width= 8.5 cm]{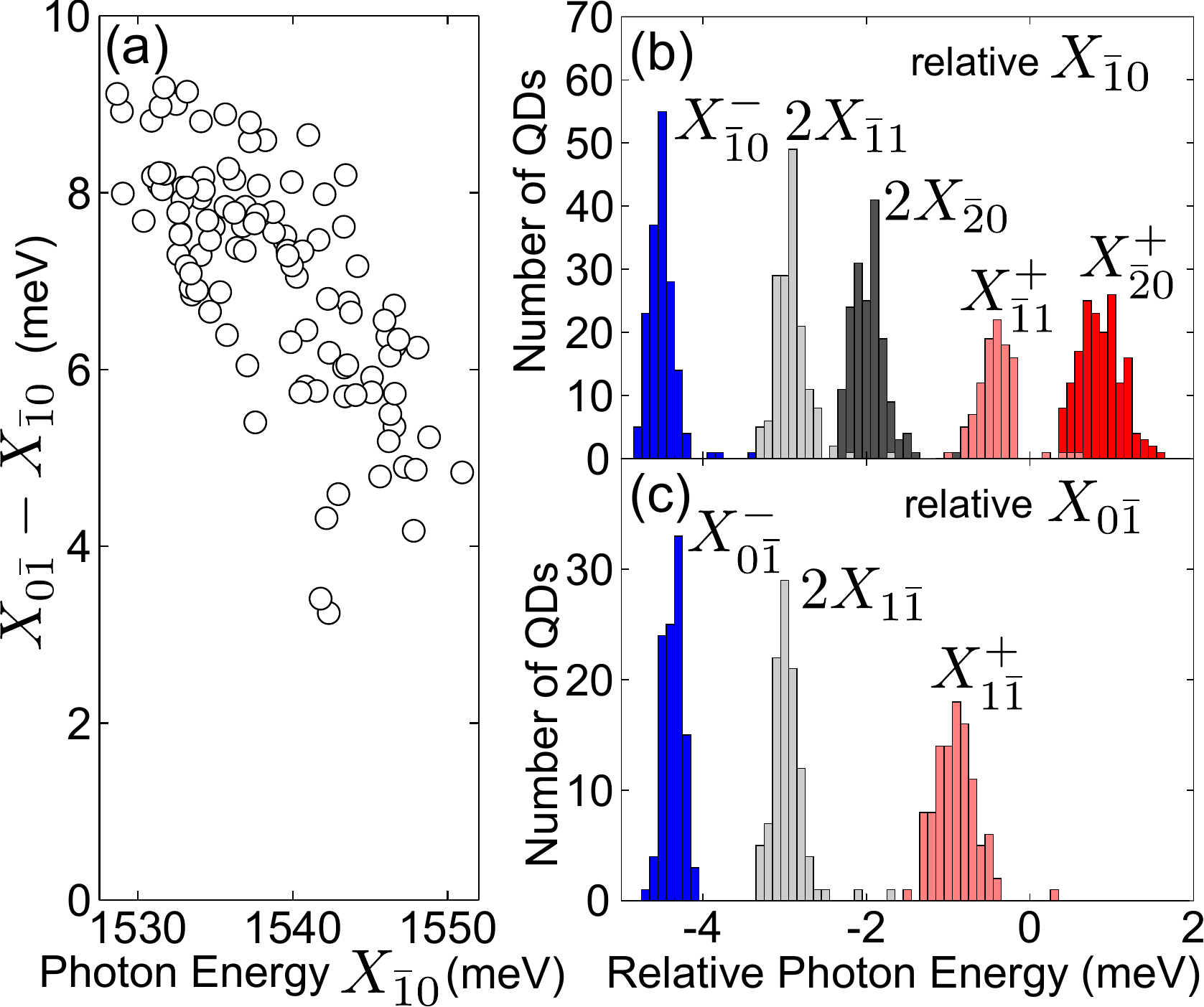}
\caption{(Color online) (a) Energy difference between the lh-like and hh-like single excitons versus the transition energy of the hh-like single exciton. (b) Histograms of hh-like transition energies relative to that of hh-like single exciton. (c) Histograms of lh-like transition energies relative to that of the lh-like single exciton.}
\label{ehist}
\end{figure}

The spectral position of the single exciton emission varies slightly from dot to dot due to existing fluctuations in the confinement potential. The variation is relatively large in the investigated sample, and is characterized by a standard deviation of $\sim$5 meV at an average $X_{\bar{1}0}$ emission energy of 1537 meV; this should be compared with later generations of pyramidal QDs that exhibit a statistical broadening as small as 1 meV \cite{Small.6.1268}. The energy separation between the lh-like and hh-like transitions $X_{0\bar{1}}$ and $X_{\bar{1}0}$ is found to be 7.1$\pm$1.3 meV, and it is clearly correlated with the emission energy of $X_{\bar{1}0}$, as seen in Fig. ~\ref{ehist} (a). This dependence originates from the stronger localization of the ground hole state $h_1$, which is thus more sensitive  to local potential fluctuations than the excited state $h_2$, causing the energy spacing between the hole levels to vary with the emission energy. In particular, the light mass of $h_2$ makes its wave function delocalized in the $z$-direction with significant leakage into the VQWR.

The actual energy spacing between $h_2$ and $h_1$ can be determined with aid of the $X^+_{\bar{1}1}$ and $X^+_{1\bar{1}}$ transitions. As illustrated in Fig. ~\ref{dec_2X11_Xp11} (a), the difference between the corresponding lh-like and hh-like transition energies of this complex gives direct access to the energy spacing between $h_2$ and $h_1$, without any influence of Coulomb interactions. The measured average spacing is in this case 6.78$\pm$1.07 meV: it is slightly smaller than that of the single excitons; The average difference between the spacings of the single excitons ($X_{0\bar{1}}$ and $X_{\bar{1}0}$) and the actual spacing of the hole levels ($h_2$ and $h_1$) is 0.56$\pm$0.13 meV. The only distinction between $X_{0\bar{1}}$ and $X_{\bar{1}0}$ is the hole configuration. Thus, the difference of $\sim$0.6 meV simply corresponds to the difference of the exciton binding energies between $X_{10}$ and $X_{01}$. It implies that the attractive e-h Coulomb interaction is weaker for $h_2$ than for $h_1$, which is not surprising, since $h_2$ is more delocalized than $h_1$, and therefore its wave function has a smaller overlap with that of the the electron. What may be unexpected, however, is that the difference in binding energy between the hh-like and lh-like exciton is only $\sim$0.6 meV despite of the lh-character of the latter exciton state.

The binding energy ($E_b$) of a complex that involves more than a single electron or hole is defined as the energy required to hypothetically dissociate an exciton from the additional carriers. If both $h_1$ and $h_2$ levels are populated with holes, like for the trion $X^+_{11}$, two such binding energies can be defined for the dissociation of either  $X_{10}$ or $X_{01}$ from a hole in $h_2$ or $h_1$. In order to distinguish between these two binding energies, the hole of the dissociated exciton will be marked with a circumflex. Accordingly, the two binding energies given in this example will be denoted $E_b(X^+_{\hat{1}1})$ and $E_b(X^+_{1\hat{1}})$. Spectroscopically, these binding energies are obtained as the difference between the transition energies of the corresponding exciton and the complex, e.g. $E_b(X^+_{\hat{1}1})$ = $X_{\bar{1}0}$ - $X^+_{\bar{1}1}$ and $E_b(X^+_{1\hat{1}})$ = $X_{0\bar{1}}$ - $X^+_{1\bar{1}}$.

A narrow distribution was observed in the relative emission energies of the exciton complexes, as shown in the histograms of Fig. ~\ref{ehist}. The average binding energies for the hh-like complexes in this sample are $E_b(X^-_{\hat{1}0})$ = 4.50$\pm$0.17 meV, $E_b(2X_{\hat{2}0})$= 1.97$\pm$0.21 meV and $E_b(X^+_{\hat{2}0})$=-0.86$\pm$0.27 meV [see Fig. ~\ref{ehist} (a)].

There is a clear dependence of the binding energy on the net charge of the complex \cite{Lelong1996819}. This dependence is common for various QD systems, and it is very well understood as a consequence of the effective masses of the holes being larger than the effective mass of the electron, which implies a stronger spatial localization of the holes compared to the electron. The repulsive direct h-h Coulomb interaction ($V_{hh}$) is therefore stronger than the attractive e-h interaction ($V_{eh}$) which, in turn, is stronger than the repulsive e-e interaction ($V_{ee}$) \cite{PhysRevB.58.16221}. The simplest model of an exciton complex accounts for a single configuration of the electrons and holes on the QD electronic levels neglecting Coulomb correlation effects, and assumes the strong confinement limit for the carriers. In such a simple model, the binding energies of the complexes are expressed to the first order of perturbation theory as: 

 \begin{eqnarray}
 E_b(X^-_{\hat{1}0}) & = & V_{eh} - V_{ee}  \nonumber \\
 E_b(2X_{\hat{2}0})  & = & 2V_{eh} - V_{hh}-V_{ee} \\
 E_b(X^+_{\hat{2}0}) & = & V_{eh} - V_{hh}\nonumber
 \end{eqnarray}

It is clear from the energy hierarchy $V_{ee}<V_{eh}<V_{hh}$ that $E_b(X^-_{\hat{1}0})$ is positive and $E_b(X^+_{\hat{2}0})$ is negative, while the value of $E_b(2X_{\hat{2}0})$ must lie in between these two trion binding energies. The Coulomb correlation effects, which are not accounted for in these expressions, are known to enhance the binding energy of the exciton \cite{PhysRevB.37.8763} and would also enhance the binding energies of all exciton complexes \cite{PhysRevB.64.165301}, possibly shifting $E_b(X^+_{\hat{2}0})$ to positive values. 

\begin{figure}[top] 
\includegraphics[width= 8.5 cm]{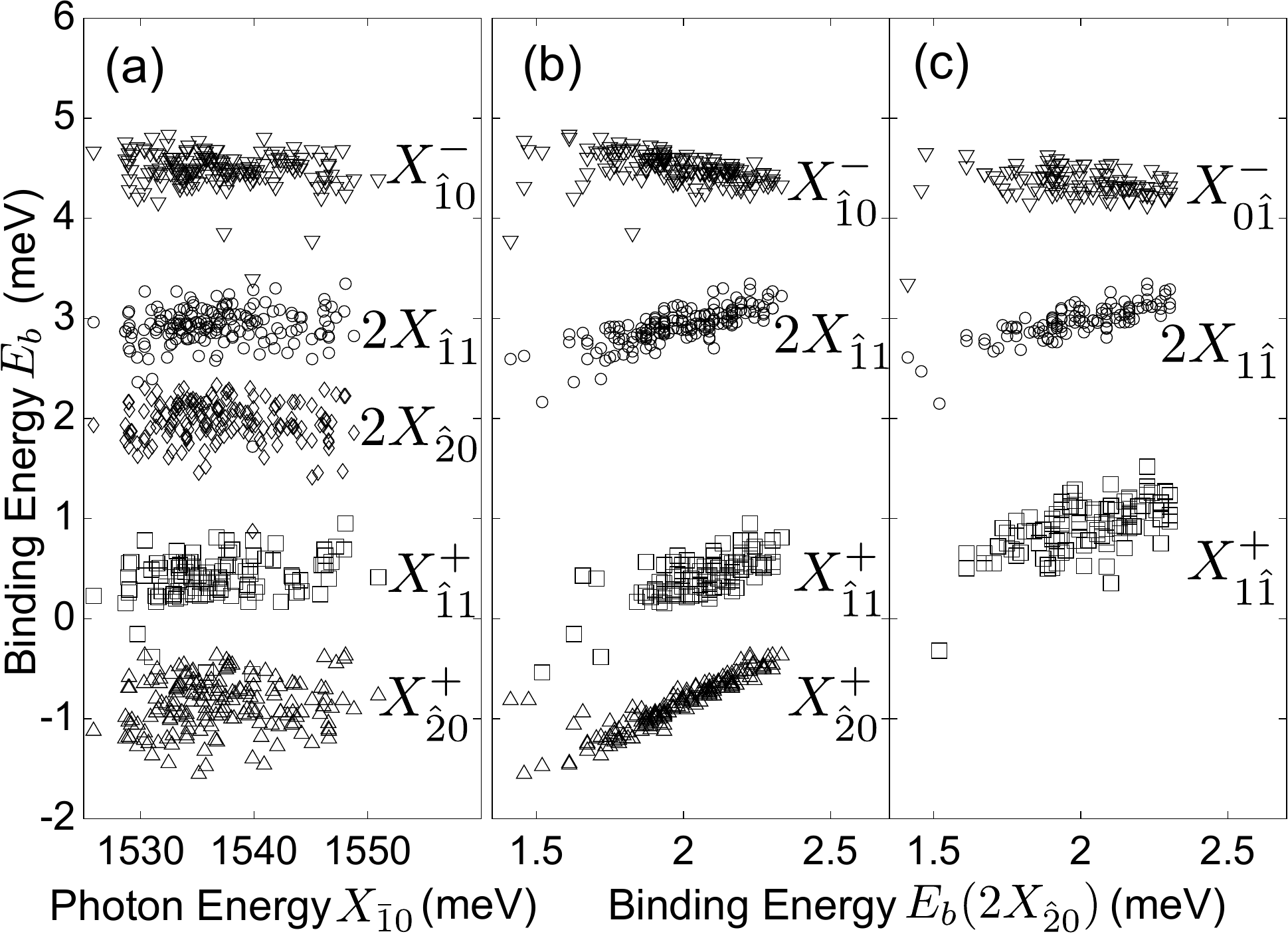}
\caption{(a) Binding energies of exciton complexes plotted versus the transition energy of the hh-like single exciton. (b) and (c) Binding energies of of exciton complexes plotted versus the binding energy of the hh-like biexciton.}
\label{becorr}
\end{figure}

A dot-to-dot variation of the Coulomb interaction energies can be caused by potential fluctuations related to a small randomness of dot size and composition. For example, a dot with a higher In-composition forms a deeper confinement potential leading not only to lower emission energy, but also to stronger Coulomb interactions with larger magnitudes of $V_{eh}$, $V_{hh}$ and $V_{ee}$. A correlation between the absolute emission energy and the binding energies of the complexes may therefore be expected. This was investigated by Baier \cite{Baier-thesis} in similar pyramidal QDs, and, surprisingly, such correlations could not be found. The absence of a correlation is also confirmed for the samples of our study. The results are displayed in Fig. ~\ref{becorr} (a), where the binding energies are plotted versus the emission energy $X_{\bar{1}0}$. Thus, we must conclude that competing mechanisms are active, mechanisms for which a deeper confinement is associated with a higher emission energy. A plausible competing mechanism is composition fluctuations in the AlGaAs barriers surrounding the dot. In this case an increased Al-composition of the barrier, for example, would also lead to deeper confinement potential and stronger Coulomb interactions, but instead to a higher emission energy. This competing mechanism is not possible for a binary barrier such a GaAs. Correlations with the emission energy $X_{\bar{1}0}$ have indeed been reported for the biexciton and negative trion binding energies in a recent study of pyramidal InGaAs QDs in pure GaAs barriers \cite{APL.101.191101}.  


The changes of the binding energies $\Delta E_b$ due to variations of the Coulomb interaction energies ($\Delta V_{eh}$,  $\Delta V_{ee}$ and $\Delta V_{hh}$) can be formulated analogously to Eq. 1.
 \begin{eqnarray}
 \Delta E_b(X^-_{\hat{1}0}) & = & \Delta V_{eh} - \Delta V_{ee}  \nonumber \\
 \Delta E_b(2X_{\hat{2}0})  & = & 2\Delta V_{eh} - \Delta V_{hh}-\Delta V_{ee} \\
 \Delta E_b(X^+_{\hat{2}0}) & = & \Delta V_{eh} - \Delta V_{hh}\nonumber
 \end{eqnarray}
The signs of these energy variations $\Delta V_{eh}$, $\Delta V_{hh}$ and $\Delta V_{ee}$ are all equal since the corresponding interaction energies $V_{eh}$, $V_{hh}$ and $V_{ee}$ are expected to be enhanced (or reduced) together under dot-to-dot potential variations. If the energy variations also obey the inequalities $|\Delta V_{ee}|<|\Delta V_{eh}|<|\Delta V_{hh}|$, then it follows from Eq. 2 that an enhancement (reduction) of all the Coulomb interaction energies corresponds to $\Delta E_b(X^-_{\hat{1}0})>0$ ($\Delta E_b(X^-_{\hat{1}0})<0$) and $\Delta E_b(X^+_{\hat{2}0})<0$ ($\Delta E_b(X^+_{\hat{2}0})>0)$. Hence, in this case the variations of the positive and negative trion binding energies are anticorrelated. Theoretically, such anticorrelation was predicted in model calculations of these charged exciton complexes in spherical QDs \cite{PhysRevB.66.165331}. The anticorrelation between the energy shifts of the positive and negative trions is observed for all the samples in our study, see ~\ref{becorr} (b), and in an earlier study on different samples \cite{Baier-thesis}. Moreover, analogous correlation patterns can be observed also for the exciton complexes involving lh-like holes, as shown in Fig. ~\ref{becorr} (c), which is consistent with the assignments of these complexes given in the previous section. Note that the negative slope of $E_b(X^-_{0\hat{1}})$ is smaller than for $E_b(X^-_{\hat{1}0})$ in Fig. ~\ref{becorr}, indicating that the magnitudes of $\Delta V_{eh}$ and $\Delta V_{ee}$ are more similar for $h_2$ than for $h_1$.  In addition, we find a strong correlation between the positive trion $E_b(X^+_{\hat{2}0})$ and the biexciton $E_b(2X_{\hat{2}0})$, implying that $|\Delta V_{ee}| + |\Delta V_{hh}| > 2 |\Delta V_{eh}|$ for the model discussed here. This relation can be reformulated as $|\Delta V_{eh} - \Delta V_{ee}| < |\Delta V_{eh} -|\Delta V_{hh}|$ and using Eq. 2 as $| \Delta E_b(X^-_{\hat{1}0})| < |\Delta E_b(X^+_{\hat{2}0})|$. Thus, this result explains the fact that the statistical variation of the binding energies of these complexes exhibit a dependence on the net charge of the complex, with smallest spread of 0.17 meV for $E_b(X^-_{\hat{1}0})$ and largest spread of 0.27 meV for $E_b(X^+_{\hat{2}0})$ [see Figs. \ref{ehist} (b) and (c)].

\section{Excitonic fine structure}
The rigorous and complete spectral identification of the dominating exciton complexes ($X_{10}$, $X^-_{10}$, $X^+_{20}$, $2X_{20}$, $X_{01}$, $X^-_{01}$, $X^+_{11}$, $2X_{11}$ and $2X^+_{21}$) demonstrated in section IV enables a detailed investigation of the fine structure in their emission patterns. It is found from experiments that the fine structure of the hybrid complexes, populated with both hh- and lh-like holes, varies significantly from dot-to-dot, both in terms of number of emission lines as well as their polarization. It is well known that the excitonic fine structure caused by Coulomb exchange interactions is intimately linked with the symmetries of the involved electron and hole states and the occupation of the single-particle levels. Therefore, any analysis of the fine structure is naturally based on symmetry arguments provided by group theory using the approach described in \cite{PhysRevLett.107.127403}.

\subsection{Group theory}
\begin{figure}[top] 
\includegraphics[width= 8.5 cm]{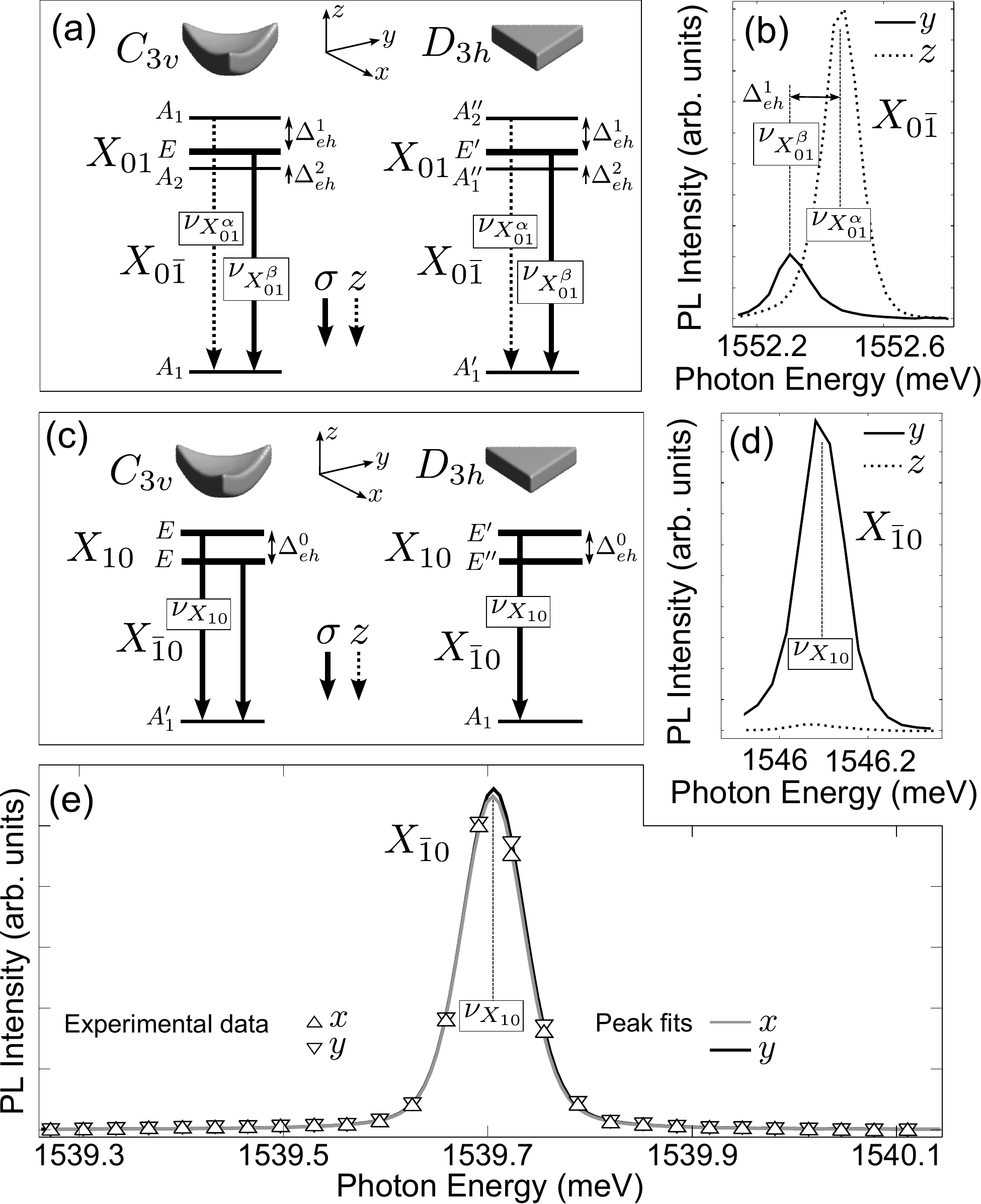}
\caption{Group theory derived decay schemes of $X_{01}$ and $X_{10}$ under (a) $C_{3v}$ and (d) $D_{3h}$. Doubly degenerate (non-degenerate) levels are shown with thick (thin) horizontal lines. Transitions with isotropic polarization in the $xy$-plane are represented by thick vertical lines, and $z$-polarized transitions are represented by dotted vertical lines. Experimental $\mu$PL data of $X_{0\bar{1}}$ (b) and $X_{\bar{1}0}$ acquired in the side-view geometry. Solid (dotted) curves correspond to $y$-polarized ($z$-polarized) PL. Black thin vertical dotted lines aid the labeling of individual emission lines. Top-view $\mu$PL data $X_{0\bar{1}}$ (e) shown as up (down) pointing triangles along with gray (black) curve fits for $x$-polarized ($y$-polarized) detection.}
\label{dec_X01_X10}
\end{figure}

The investigated zincblende pyramidal QDs grown along [111] ideally possess three symmetry planes, corresponding to point group $C_{3v}$ \cite{PhysRevB.81.161307}. Hence, the pyramidal QDs can possess a symmetry higher than conventional QDs grown on (001)-planes, for which the zincblende crystal is compatible only with two symmetry planes ($C_{2v}$). The quantum states, which in general have symmetries different from that of the actual QD, are labeled according to the irreducible representations of the point group. On the basis of a few simple arguments \cite{PhysRevLett.107.127403} it can be shown that the electrons in the ground state of a $C_{3v}$ QD are restricted to the double group representation $E_{1/2}$, using the Mulliken notation of Ref. \onlinecite{Altmann}, while two types of holes can exist, labeled either $E_{1/2}$ or $E_{3/2}$ (strictly speaking $^1E_{3/2}$ + $^2E_{3/2}$). In the strong confinement regime, the corresponding labels of the excitonic states are obtained  simply by label multiplication of the involved electron and hole. Using available multiplication tables for $C_{3v}$, the quantum states of the two possible types of single excitons are obtained \cite{Altmann}: $E_{1/2} \times E_{1/2} = A_1 + A_2 + E$ for type 1 and $E_{1/2} \times E_{3/2} = E + E$ for type 2. The polarized optical transition between the initial excitonic state and the final state of an empty QD (which is invariant to any symmetry operations of the point group and therefore labeled $A_1$), can be examined in the dipole-approximation by the Wigner-Eckart theorem. The dipole operator itself transforms according to conventional vectors (in $C_{3v}$ labeled $E$ for $x$- and $y$-polarization, and $A_1$ for $z$-polarization) and the resulting optical decay schemes are shown in Fig. ~\ref{dec_X01_X10} (a and c). Eventual energy spacings between the excitonic states are solely caused by the electron-hole exchange interaction ($\Delta_{eh}$). Note that the energy order of the states is not determined by symmetry arguments, but the order in Fig. ~\ref{dec_X01_X10} is chosen to be consistent with experiments. In Fig. ~\ref{dec_X01_X10} we also display the group theoretical prediction corresponding to the higher symmetry group $D_{3h}$, obtained from $C_{3v}$ by adding the symmetry operation of a symmetry plane perpendicular to those of $C_{3v}$. In such case, the ground state electrons is associated to $E_{5/2}$ and the two hole states to $E_{1/2}$ and $E_{3/2}$ \footnote{Note that in Refs. \cite{PhysRevB.81.161307} and \cite{PhysRevLett.107.127403}, the irreducible representations $E_{1/2}$  and $E_{5/2}$ were interchanged. This does not lead to different physical predictions.}. 

\subsection{Spectral analysis}
From the Wigner-Eckart theorem, we predict that the spectrum of a type 1 exciton consists of one component \emph{isotropically polarized} in the $xy$-plane ($\sigma$-polarized) as well as another $z$-polarized component, see Fig. ~\ref{dec_X01_X10} (a). This is in perfect agreement with the experimental spectra of the lh-like exciton $X_{01}$ shown in Fig. ~\ref{dec_X01_X10} (b). The energy splitting between the two optically active states of $X_{01}$ can be determined directly from the spectrum $\Delta^1_{eh}$ = 155 $\mu$eV, but the splitting with the dark state, $\Delta^2_{eh}$, cannot be determined by any direct measurements of $X_{0\bar{1}}$. 
  
An exciton of type 2 is predicted to be optically active only with $\sigma$-polarization, like the hh-like exciton $X_{10}$ shown in Fig. ~\ref{dec_X01_X10} (c). However, the theory predicts two optically active states for $X_{\bar{1}0}$ while the experimental spectra of $X_{10}$ reveal only a single emission line, see Fig. ~\ref{dec_X01_X10} (d). This suggests either that the splitting $\Delta^0_{eh}$ between the energy levels is too small to be resolved, or that the intensity of one of the components is too weak to be detected. The absence of a second $\sigma$-polarized emission line of $X_{\bar{1}0}$ is further supported by the data of another dot measured in top-view configuration, giving access to both the $x$- and $y$-polarization shown in Fig. ~\ref{dec_X01_X10} (e). The existence of a single emission line is consistent with the predicted polarization isotropy within the $xy$-plane.

In order to explain the discrepancy between the theoretically derived decay schemes for $C_{3v}$ and the actual PL spectrum of $X_{10}$, the concept of symmetry elevation was introduced in our previous works \cite{PhysRevB.81.161307, PhysRevLett.107.127403}. It was argued that an additional horizontal symmetry plane can be assumed, causing an approximate elevation of the symmetry from $C_{3v}$ to $D_{3h}$. The corresponding decay schemes for $D_{3h}$ are also shown in Figs. ~\ref{dec_X01_X10} (a) and ~\ref{dec_X01_X10} (c).  In the case of type I exciton, the fine structure splitting pattern of the exciton is unchanged, but one of the optically active state of $X_{0\bar{1}}$ becomes a dark state when the symmetry is elevated to $D_{3h}$; as a result, there is a single optical transition in the emission spectrum of $X_{0\bar{1}}$, in agreement with the experimental data reported in Figs.  ~\ref{dec_X01_X10} (d) and ~\ref{dec_X01_X10} (e).

\begin{figure}[top] 
\includegraphics[width= 8.5 cm]{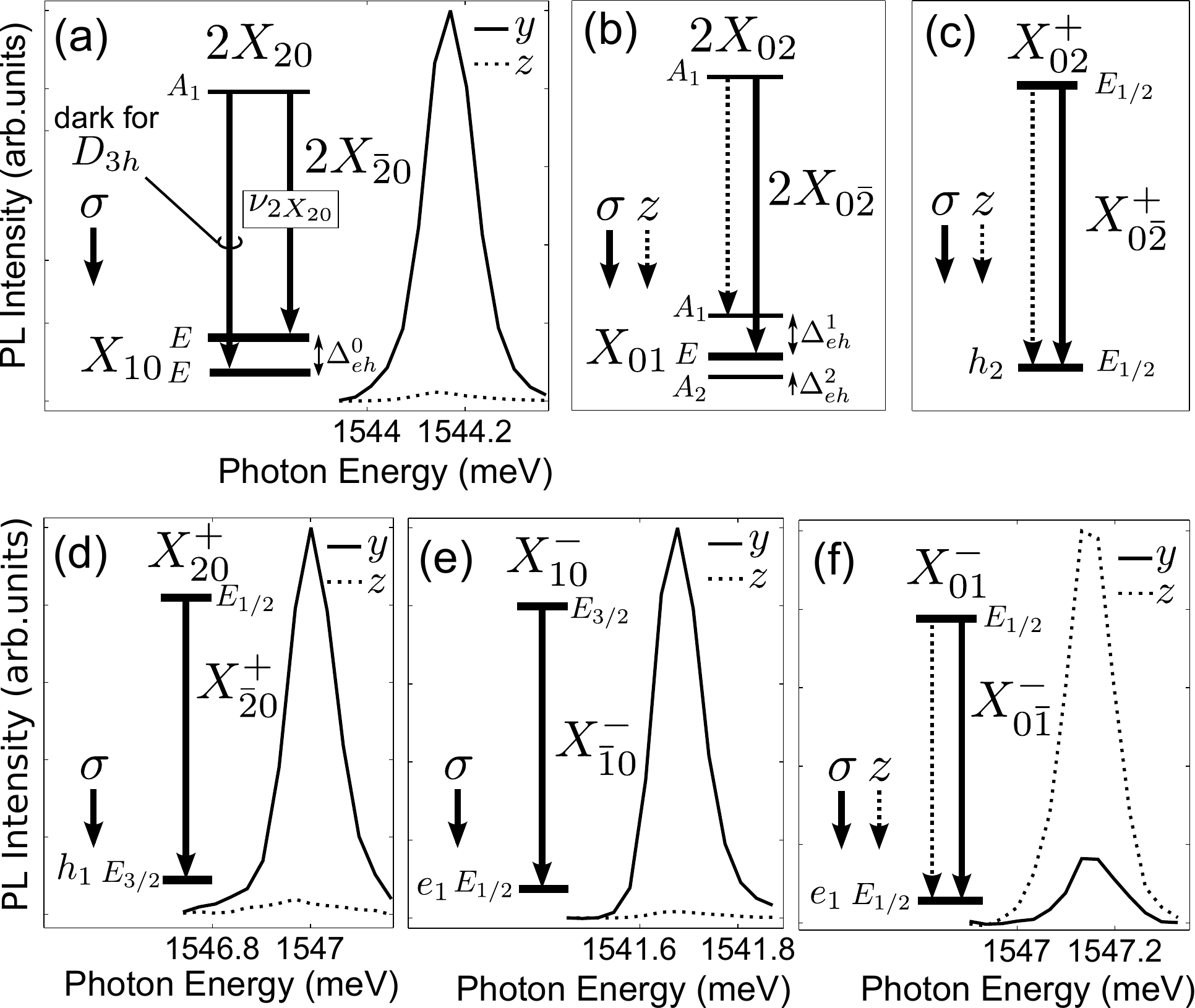}
\caption{Group theory derived decay schemes under $C_{3v}$ and experimental $\mu$PL data acquired in the side-view geometry of indicated exciton complexes. Doubly degenerate (non-degenerate) levels are shown with thick (thin) horizontal lines. Transitions with isotropic polarization in the $xy$-plane are represented by thick vertical lines, and $z$-polarized transitions are represented by dotted vertical lines. Solid (dotted) curves correspond to $y$-polarized ($z$-polarized) PL.}
\label{dec_2X20}
\end{figure}

For exciton complexes involving more than one electron-hole pair, the irreducible representations of its quantum states are obtained by the product of all involved electrons and holes. For a completely filled electron (hole) level, the Pauli exclusion principle restricts the product of electrons (holes) representations to the invariant irreducible representation $A_1$ ($A_1'$) in $C_{3v}$ ($D_{3h}$). Note that the single particle energy levels are always doubly degenerate because of the Kramers theorem.

The biexciton $2X_{20}$ consists of one filled electron level and one filled hole level, the total product is therefore also invariant $A_1 \times A_1 = A_1$, like an empty QD. Consequently, the fine structure splitting pattern of this biexciton mimics the pattern of the exciton, but with a reversed energy order. The decay scheme derived for $2X_{20}$ under $C_{3v}$ is shown in Fig. ~\ref{dec_2X20} (a); the experimental spectra, only reveal a single emission line, which is in contrast to the prediction of two lines in the case of the point group $C_{3v}$. Analogously to our previous result for the exciton $X_{10}$, one of the transitions of $2X_{\bar{2}0}$ becomes optically inactive, providing further evidence of elevated symmetry $D_{3h}$ for these states. The corresponding lh-like biexciton $2X_{0\bar{2}}$ has not been resolved experimentally, but its spectral pattern is also predicted to mimic the reversed pattern of the single exciton $X_{0\bar{1}}$, which is identical for both $C_{3v}$ and $D_{3h}$, see Fig. ~\ref{dec_2X20} (b).

Several trions are formed from one filled electron or hole level, plus an additional carrier of the opposite charge. The final state of an optical transition is then a single carrier. In this case, the irreducible representation of the initial state is determined solely by the additional carrier, while the label of the final state is determined solely by the remaining carrier. As the single particle states are Kramers degenerate (in absence of external magnetic fields), single emission lines are predicted for such trions, see Figs. ~\ref{dec_2X20} (c) to (f). Moreover, the hh-like trions are predicted to emit light only with $x$- and $y$-polarization, in agreement with the experimental data also shown in in Figs. ~\ref{dec_2X20} (d) and ~\ref{dec_2X20} (e), while the lh-like trions are predicted to emit light polarized both in the $xy$-plane as well as in the $z$-direction, as confirmed by the spectra of the negative trion $X^-_{01}$ in Fig. ~\ref{dec_2X20} (f). The positive lh-like trion $X^+_{0\bar{2}}$ as a single line is supported by a faint emission line in the spectra of Fig. ~\ref{pano_CE_symQD} (b) in full consistency with the derived pattern of Fig. ~\ref{dec_2X20} (c), but it needs to be discussed in more detail, which will be done below. All the spectral patterns derived for different trion species in Figs. ~\ref{dec_2X20} (c) to (f) are identical for both $C_{3v}$ and $D_{3h}$.

Symmetry elevation was similarly found for the complicated hybrid exciton complexes involving holes of different characters \cite{PhysRevLett.107.127403}, e.g. the biexciton $2X_{11}$. In this case, the Pauli exclusion applies only to the two electrons, resulting in the invariant product labeled $A_1$ for point group $C_{3v}$. The two holes, on the other hand, occupy unfilled levels yielding the product $E_{1/2} \times E_{3/2} = E + E$. Finally, the biexcitonic states are obtained as the product of these intermediate factors, $A_1 \times (E + E) = E + E$. Thus, any energy spacing between the two $E$-states of $2X_{11}$ originates from the h-h exchange interaction ($\Delta_{hh}$). The final states of the transitions $2X_{1\bar{1}}$ and $2X_{\bar{1}1}$ are the two single excitons $X_{10}$ and $X_{01}$, respectively.

\begin{figure}[top] 
\includegraphics[width= \columnwidth]{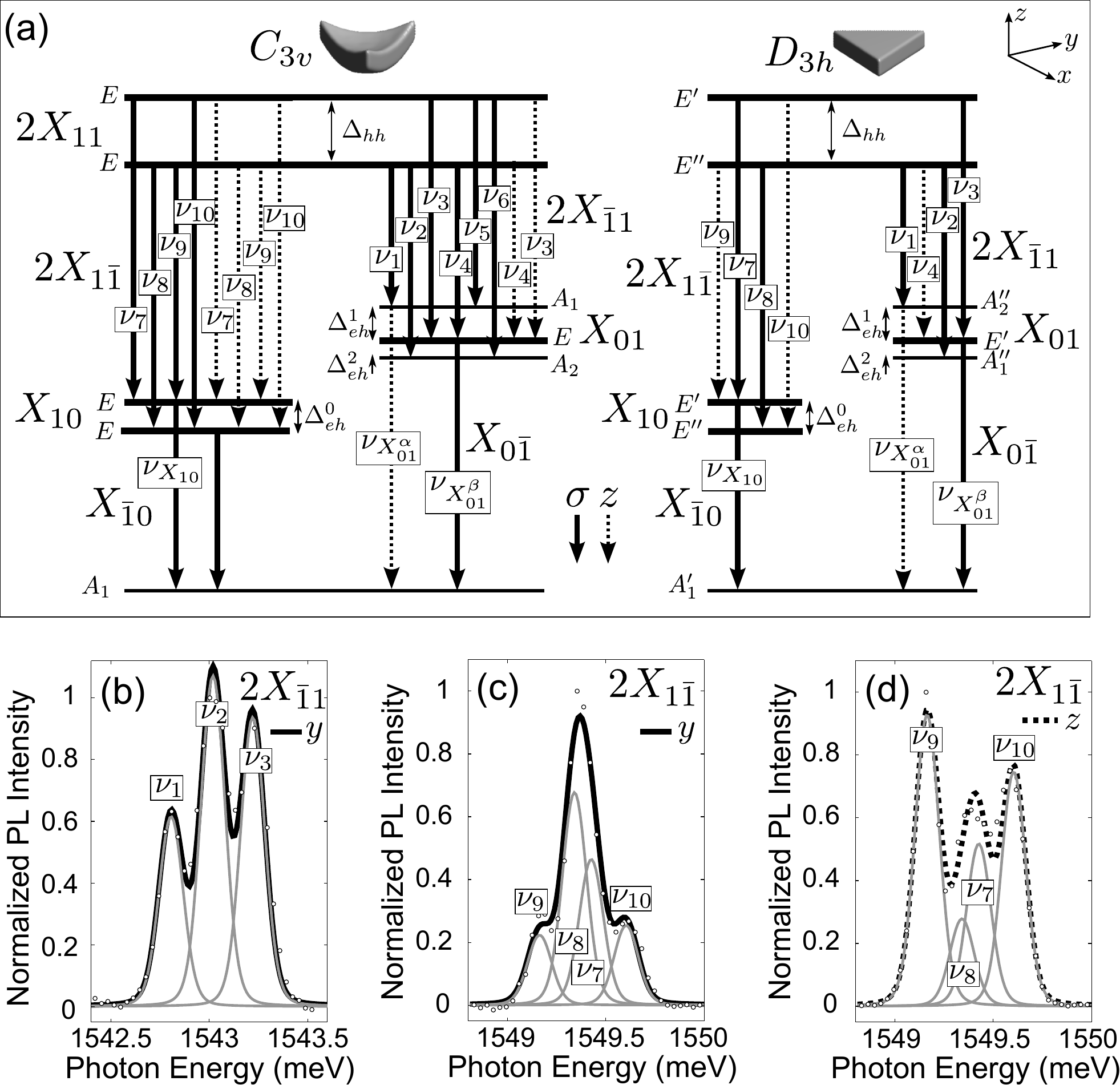}
\caption{\label{Fig. 14} (a) Group theory derived decay schemes of $2X_{11}$ under $C_{3v}$ and $D_{3h}$. Doubly degenerate (non-degenerate) levels are shown with thick (thin) horizontal lines. Transitions with isotropic polarization in the $xy$-plane are represented by thick vertical lines, and $z$-polarized transitions are represented by dotted vertical lines. (b) to (d) Experimental $\mu$PL data of indicated exciton complexes acquired in the side-view geometry shown as circles along with peak fits shown as thick solid or dotted curves for $y$-Êand $z$-polarized data, respectively. The individual Voigt peaks used in the fitting have identical widths and they are shown with grey solid lines below each curve. }
\label{dec_2X11}
\end{figure}

The full decay schemes of $2X_{11}$, derived for both $C_{3v}$ and $D_{3h}$, are shown in Fig. ~\ref{dec_2X11} (a). In this case, symmetry elevation considerably simplifies the optical spectrum. For instance, the six $\sigma$-polarized transitions $\nu_1$ to $\nu_6$ predicted for $2X_{\bar{1}1}$ under $C_{3v}$ reduce to three transitions $\nu_1$ to $\nu_3$ under $D_{3h}$. The actual $y$-polarized PL spectrum of $2X_{\bar{1}1}$ can indeed be satisfactorily explained by merely three emission lines, as shown by a peak fit in Fig. ~\ref{dec_2X11} (b). However, precise measurements performed in the top-view geometry evidence also weak features related to $\nu_4$ to $\nu_6$ \cite{PhysRevLett.107.127403}. Note that $z$-polarized components $\nu_3$ and $\nu_4$ of $2X_{\bar{1}1}$ are expected to be very weak due to the strong hh-like character of the recombining hole, hence the fact that no appreciable $z$-polarized intensity is detected in the experiments for these transitions is well understood. For the lh-like transitions $2X_{1\bar{1}}$, on the other hand, the $z$-polarized components are expected to dominate. A symmetry analysis based on $C_{3v}$ predicts four transitions $\nu_7$ to $\nu_{10}$ optically active with both $\sigma$- and $z$-polarization, while this pattern simplifies under $D_{3h}$ where only two ($\nu_7$ and $\nu_8$) are active with $\sigma$-polarization and the other two ($\nu_9$ and $\nu_{10}$) are active with $z$-polarization. The experiments reveal two transitions ($\nu_7$ and $\nu_8$) dominating with $y$-polarization [see peak fits in Fig. ~\ref{dec_2X11} (c)], and two other transitions ($\nu_9$ and $\nu_{10}$) dominating with $z$-polarization [see peak fits in Fig. ~\ref{dec_2X11} (d)]. However, all four emission lines $\nu_7$ to $\nu_{10}$ of $2X_{1\bar{1}}$ are active with both $y$- and $z$-polarization. Thus, the transitions $\nu_7$ to $\nu_{10}$ are all probing the actual $C_{3v}$ symmetry of the dot, but their relative intensities can be understood on the basis of an approximate elevation of symmetry towards $D_{3h}$.

Having identified all the relevant emission lines of $X_{10}$, $X_{01}$ and $2X_{11}$ in the PL spectra, experimental values of the fine structure energies can conveniently be extracted with the aid of the diagrams in Fig. ~\ref{dec_2X11} (a). It is found that for the dot presented in Figs. ~\ref{dec_X01_X10} (b-c) and ~\ref{dec_2X11} (b-d), $\Delta_{eh}^0 = 172 \mu$eV, $\Delta_{eh}^1 = 155 \mu$eV, $\Delta_{eh}^2 = 62 \mu$eV and $\Delta_{hh} = 265 \mu$eV. The same analysis performed on other QDs in the same sample give very similar values. A splitting between the two energy levels of $X_{10}$ is determined within a range of 150-180 $\mu$eV, which is sufficiently large to be easily resolvable in the PL spectra. However, no additional spectral feature can be observed in a range of $\pm$300 $\mu$eV  from the single emission line of $X_{10}$ as shown in Fig. ~\ref{dec_X01_X10} (e), thereby confirming the existence of dark states resulting from a symmetry elevation towards $D_{3h}$.

In the following, we will describe the transitions related to the positively charged biexciton $2X^+_{21}$. The electron level and the first hole level are  both filled for this complex. Hence, the Pauli exclusion principle leads to invariant e-e and h-h states labeled $A_1$ under $C_{3v}$. Thus, the symmetry of the quantum states of $2X^+_{21}$ is merely determined by the additional lh-like hole state labeled $E_{1/2}$, since $A_1 \times A_1  \times E_{1/2} = E_{1/2}$. The charged biexciton $2X^+_{21}$ decays radiatively either via the conventional trion $X^+_{20}$ or the excited trion $X^+_{11}$. In the case of $X^+_{20}$ the hole level is also filled, resulting in an invariant h-h state $A_1$. The quantum state of this complex is therefore entirely determined by the additional electron state labeled $E_{1/2}$. For $X^+_{11}$, on the contrary, the holes form two $E$-states split by h-h exchange interaction, analogous to $2X_{11}$, which, when multiplied with the single electron, yield the total product $E_{1/2} \times (E + E) = E_{3/2} + E_{1/2} + E_{3/2} + E_{1/2}$. Hence, unlike $2X_{11}$, the states of $X^+_{11}$ are additionally split by e-h exchange interactions. The trions finally decay into single hole states $h_2$ or $h_1$, labeled $E_{1/2}$ and $E_{3/2}$, respectively.

\begin{figure}[top] 
\includegraphics[width= \columnwidth]{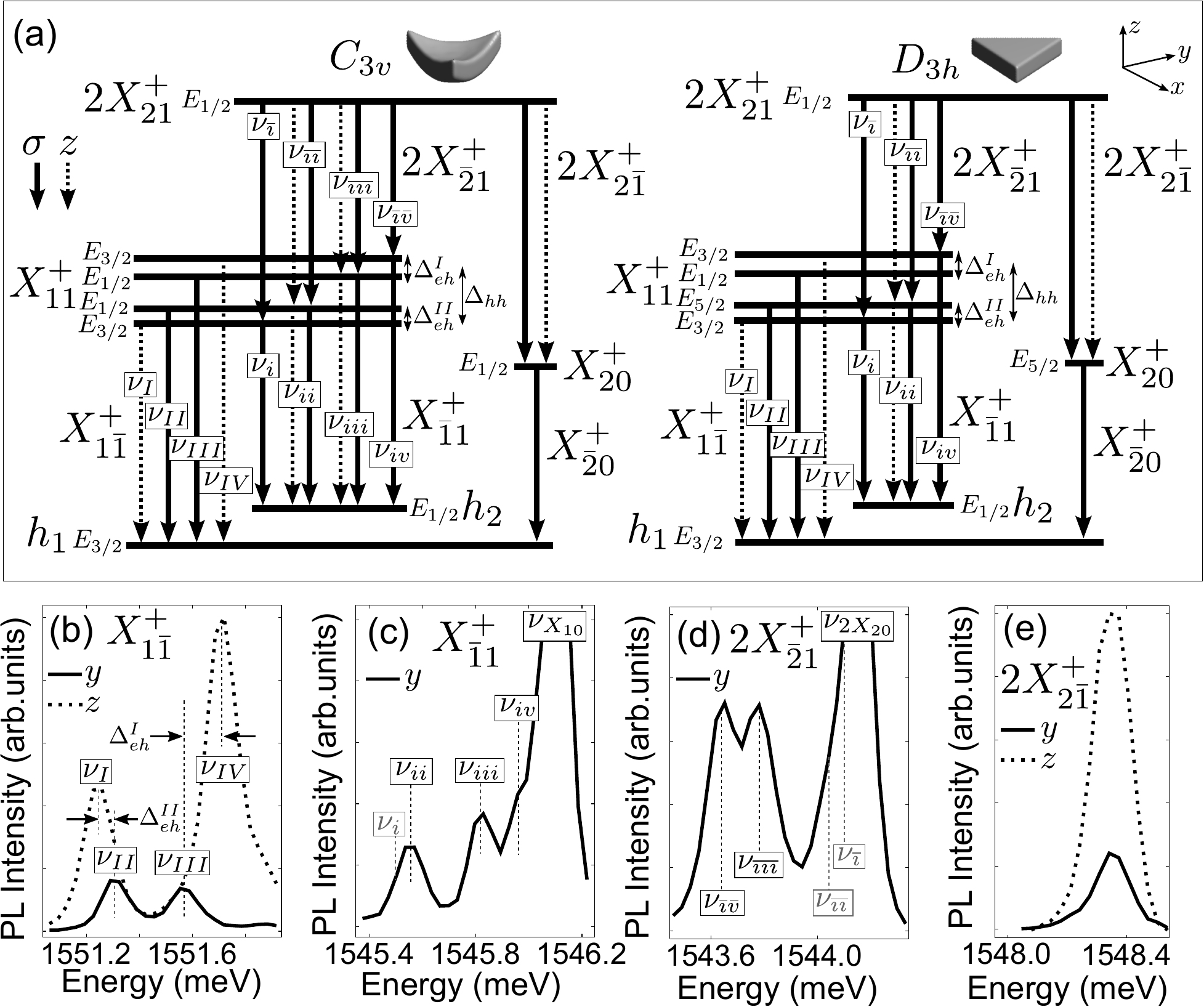}
\caption{ (a) Group theory derived decay schemes of $2X^+_{21}$ under $C_{3v}$ and $D_{3h}$. Doubly degenerate levels are shown with thick  horizontal lines. Transitions with isotropic polarization in the $xy$-plane are represented by thick vertical lines, and $z$-polarized transitions are represented by dotted vertical lines. (b) to (e) Experimental $\mu$PL spectra of indicated exciton complexes acquired in side-view geometry at $T$ = 27 K. Solid (dotted) curves correspond to $y$-polarized ($z$-polarized) PL. Black thin vertical dotted lines aid the labeling of individual emission lines. Gray vertical dotten lines labels indicate consistently predicted energy positions. The strong and overlapping intensities from $X_{10}$ and $2X_{20}$ in (c) and (d) are cropped vertically.}
\label{dec_2X21p}
\end{figure}

The complete decay schemes of $2X^+_{21}$ for both $C_{3v}$ and $D_{3h}$ are depicted in Fig. ~\ref{dec_2X21p} (a). The optical transition patterns of $X^+_{1\bar{1}}$ are identical for $C_{3v}$ and $D_{3h}$, featuring two components ($\nu_{II}$ and $\nu_{III}$) active with $\sigma$-polarized light, and two other transitions ($\nu_{I}$ and $\nu_{IV}$) active with $z$-polarized light. Thus, the spectrum of $X^+_{1\bar{1}}$ is unaffected by symmetry elevation, and the strict polarization selection rules are entirely obeyed by the experimental data shown in Fig. ~\ref{dec_2X21p} (b). The corresponding hh-like spectrum $X^+_{\bar{1}1}$, on the other hand, is predicted to be different between $C_{3v}$ and $D_{3h}$, with one of the transitions, $\nu_{iii}$, becoming optically inactive under $D_{3h}$. The experimental spectrum of $X^+_{\bar{1}1}$ shown in Fig. ~\ref{dec_2X21p} (c) overlaps partly with the strong emission (saturated) from the single exciton $\nu_{X_{10}}$, and, due to the hh-like character of the recombing hole, none of the predicted vertically polarized components can be observed. Nevertheless, the $\nu_{iii}$ transition is clearly resolved with sufficient intensity in these experiments. We conclude from this that the $\nu_{iii}$ transition is a very sensitive probe of the actual $C_{3v}$ symmetry. This conclusion also holds for the transitions related to $2X^+_{\bar{2}1}$, where the corresponding transition $\nu_{\bar{\imath}\bar{\imath}\bar{\imath}}$ becomes dark under $D_{3h}$. In this case is the $\nu_{\bar{\imath}\bar{\imath}\bar{\imath}}$ transition is also well-resolved in the PL spectrum shown in Fig. ~\ref{dec_2X21p} (d), while two of the other transitions, $\nu_{\bar{\imath}}$ and $\nu_{\bar{\imath}\bar{\imath}}$ are concealed due to overlap with the strong emission from the biexciton $\nu_{X_{20}}$. Note that even though some transitions remain unresolved, their exact energy positions can be precisely predicted from the energy spacings between the transitions that are observed in the $X^+_{1\bar{1}}$ spectrum. These predicted but unresolved transitions are indicated with gray labels and lines in Figs. ~\ref{dec_2X21p} (c-d). An experimental indication of the unresolved components $\nu_{\bar{\imath}}$ and $\nu_{\bar{\imath}\bar{\imath}}$ is, however, evidenced by the slight asymmetry of $\nu_{X_{20}}$ in Fig. ~\ref{dec_2X21p} (d). The sole emission line of $2X^+_{2\bar{1}}$ is shown in Fig. ~\ref{dec_2X21p} (e) for completeness.

In the above examples, transitions $\nu_7$ to $\nu_{10}$ and in particular $\nu_{iii}$ as well as $\nu_{\bar{\imath}\bar{\imath}\bar{\imath}}$ are found to be sensitive probes of the actual $C_{3v}$ symmetry, highlights the importance to consider the true symmetry of the QD in order to fully understand the fine structure of the spectral patterns.

The e-h exchange interactions $\Delta_{eh}^I$ and $\Delta_{eh}^{II}$ cause the energy spacing within pairs of $y$- and $z$-polarized components of $X^+_{1\bar{1}}$. Since the transitions of $X^+_{1\bar{1}}$ consist of two components in each polarization direction, these e-h exchange energies can only be determined as indicated in Fig. ~\ref{dec_2X21p} (b) under the ad-hoc assumption that $\Delta_{eh}^{I} < \Delta_{hh}$ and $\Delta_{eh}^{II} < \Delta_{hh}$. In this case, the resulting values would be $\Delta_{eh}^I = \nu_{IV} - \nu_{III} =$ 140 $\mu$eV and $\Delta_{eh}^{II} = \nu_{II} - \nu_{I} =$ 60 $\mu$eV, while $\Delta_{hh}$ then must be in the range from 260 to 400 $\mu$eV, i.e. values comparable with the exchange interaction energies previously obtained for the single excitons and the excited biexciton ($\Delta_{eh}^0$ = 172 $\mu$eV, $\Delta_{eh}^1$ = 155 $\mu$eV, $\Delta_{eh}^2$ = 62 $\mu$eV and $\Delta_{hh}$ = 265 $\mu$eV ). The assumption made here is also fully consistent with the experimental results that are presented in the next section when symmetry breaking is taking place.

\begin{figure}[top] 
\includegraphics[width= 8.5 cm]{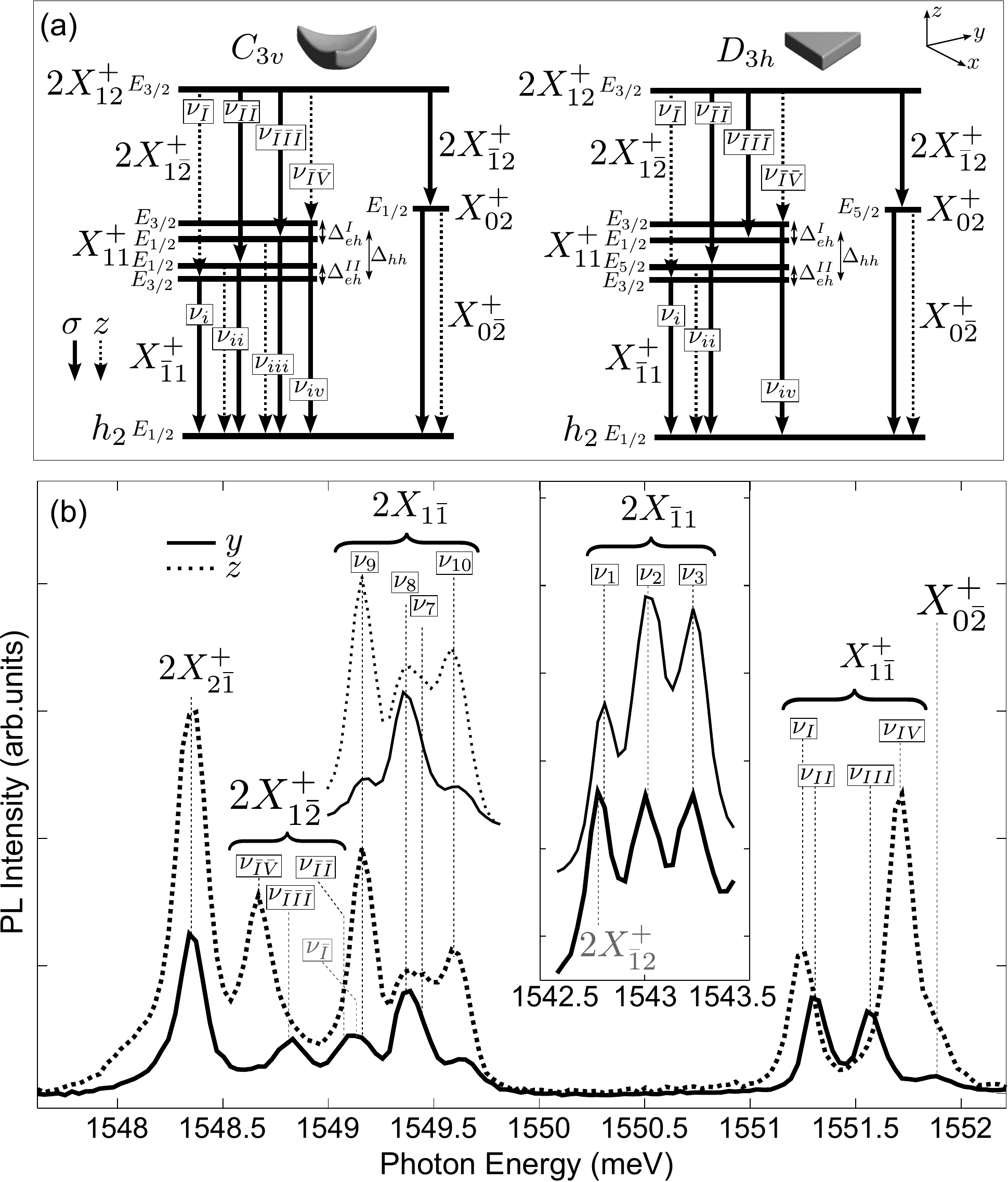}
\caption{(a) Group theory derived decay schemes of $2X^+_{12}$ under $C_{3v}$ and $D_{3h}$. Doubly degenerate levels are shown with thick horizontal lines. Transitions with isotropic polarization in the $xy$-plane are represented by thick vertical lines, and $z$-polarized transitions are represented by dotted vertical lines. (b) Experimental $\mu$PL spectra acquired in side-view geometry with exciton power of 650 nW at $T$ = 27 K. Solid (dotted) curves correspond to $y$-polarized ($z$-polarized) PL. Additional and vertically shifted spectra of $2X_{1\bar{1}}$ acquired with the lower excitation power of 500 nW are also shown with thin lines. Black thin vertical dotted lines aid the labeling of individual emission lines. Gray vertical dotted lines labels indicate consistently predicted energy positions. The inset shows spectra of $2X_{\bar{1}1}$ acquired at 650 nW and 500 nW with thick and thin lines, respectively.}
\label{dec_2X12p}
\end{figure}

The positively charged biexciton $2X^+_{21}$ is the only complex identified with lh-like emission that is in its ground state, i.e. without having any holes in higher excited states. It was shown in Fig. ~\ref{pano_CE_symQD} (b) that the appearance of $2X^+_{2\bar{1}}$ at high powers was accompanied by several weaker spectral features polarized in $z$- and $y$-direction and positioned in between $2X^+_{2\bar{1}}$ and $2X_{1\bar{1}}$. We will present a series of consistent arguments to justify the assignment of these features to the excited positively charged biexciton $2X^+_{12}$. The derived decay schemes of $2X^+_{12}$ as well as close ups of the relevant parts of the spectra are shown in Fig. ~\ref{dec_2X12p}. The theoretical fine structure of $2X^+_{1\bar{2}}$ shown in Fig. ~\ref{dec_2X12p} (a) exhibits a pattern identical with $X^+_{1\bar{1}}$, except that it is energetically reversed for $2X^+_{1\bar{2}}$. Indeed, the experimental energy difference between the $2X^+_{1\bar{2}}$ transitions $\nu_{\bar{I}\bar{V}}$ and $\nu_{\bar{I}\bar{I}\bar{I}}$ is equal to that between $\nu_{IV}$ and $\nu_{III}$ for $X^+_{1\bar{1}}$, but the spectral ordering of the two polarized transitions is reversed. Moreover, the presence of a third peak $\nu_{\bar{I}\bar{I}}$ is evidenced by the apparent broadening and slight red shift of $\nu_9$ related to $2X_{1\bar{1}}$ at high powers. Thus, it is concluded that $\nu_{\bar{I}\bar{I}}$ in Fig. ~\ref{dec_2X12p} partially overlaps with spectral line of $\nu_9$. Note that the order of the pattern is reversed in energy and that also the polarization of  $\nu_{\bar{I}\bar{I}}$ to $\nu_{\bar{I}\bar{V}}$ perfectly matches that of the transitions $\nu_{II}$ to $\nu_{IV}$ of $X^+_{1\bar{1}}$. The fourth transition $\nu_{\bar{I}}$, predicted to be $z$-polarized, is completely superimposed onto the strong $z$-polarization of $\nu_9$, and cannot be resolved, but its accurate energy position is indicated with a gray label $\nu_{\bar{I}}$ in Fig. ~\ref{dec_2X12p} (b).

There is a second recombination path $2X^+_{\bar{1}2}$ from the complex $2X^+_{12}$ that leads to the doubly excited trion $X^+_{02}$. Due to the low relaxation efficiency between the hole states that is evidenced in Fig. ~\ref{activation} (b), this trion may recombine optically before relaxation takes place. A single emission line active with both $\sigma$- and $z$-polarization is predicted for this complex [see Figs. ~\ref{dec_2X20} (c) and ~\ref{dec_2X12p} (a)]. A plausible candidate for $X^+_{0\bar{2}}$ in the spectra is the shoulder appearing on the high energy side of $\nu_{IV}$ of $X^+_{1\bar{1}}$ in Fig. ~\ref{dec_2X12p} (b) at high excitation powers. This spectral peak is most clearly resolved in the $y$-polarization, because its dominating polarization in the $z$-direction is overlapping with a strong $z$-polarized component of $X^+_{1\bar{1}}$.

With the aid of the decay diagrams in Fig. \ref{dec_2X12p} (a), it is now possible to accurately predict the energy of the corresponding hh-like transition $2X^+_{\bar{1}2}$. For this QD, the predicted energy nearly coincides with that of the $\nu_1$ transition of $2X_{\bar{1}1}$, as indicated by the gray label in the inset of Fig. \ref{dec_2X12p} (b). Thus, $2X^+_{\bar{1}2}$ remains unresolved, but it provides an explanation to the apparent change in relative intensities between $\nu_1$ and the two other transions $\nu_2$ and $\nu_3$ of the same complex at increased excitation power [see inset of Fig. \ref{dec_2X12p} (b)], as $2X^+_{12}$ appears only at excitation powers sufficiently high for hole state filling. It may also explain the apparent slight energy shift to lower frequency of the lowest $2X_{\bar{1}1}$ transition $\nu_1$ at high excitation power.

It should be noted that all the discussed effects of symmetry elevation towards $D_{3h}$ made in this section would also hold for symmetry elevation from $C_{3v}$ towards $C_{6v}$. In this case,  the electron as well as the light-hole like states are associated to $E_{1/2}$ and the heavy-hole like states to $E_{3/2}$. Thus, although the intuitive arguments for elevation towards $C_{6v}$ are less obvious than those for $D_{3h}$ \cite{PhysRevLett.107.127403}, none of the performed experiments can distinguish which type of elevation is most relevant. It may be that both elevations are relevant for certain complexes, leading to very strong and robust elevation effects, like the dark state of $X_{10}$.

To summarize this part, symmetry arguments applied to the optical selection rules and to the fine structure were sufficient to gain essential understanding of the complexity of the emission patterns exhibited by a large variety of excitonic complexes. It was demonstrated that the pyramidal quantum dots do exhibit a high $C_{3v}$ symmetry, and whereas many spectral patterns are well explained on the basis of an approximate elevation of the symmetry, some patterns and specific transitions are particularly sensitive indicators of the exact $C_{3v}$ symmetry. These results strongly highlight the importance of taking the true dot symmetry into consideration when analyzing the spectral fine structure.

\section{Signatures of symmetry breaking}

\begin{figure}[top] 
\includegraphics[width= 8 cm]{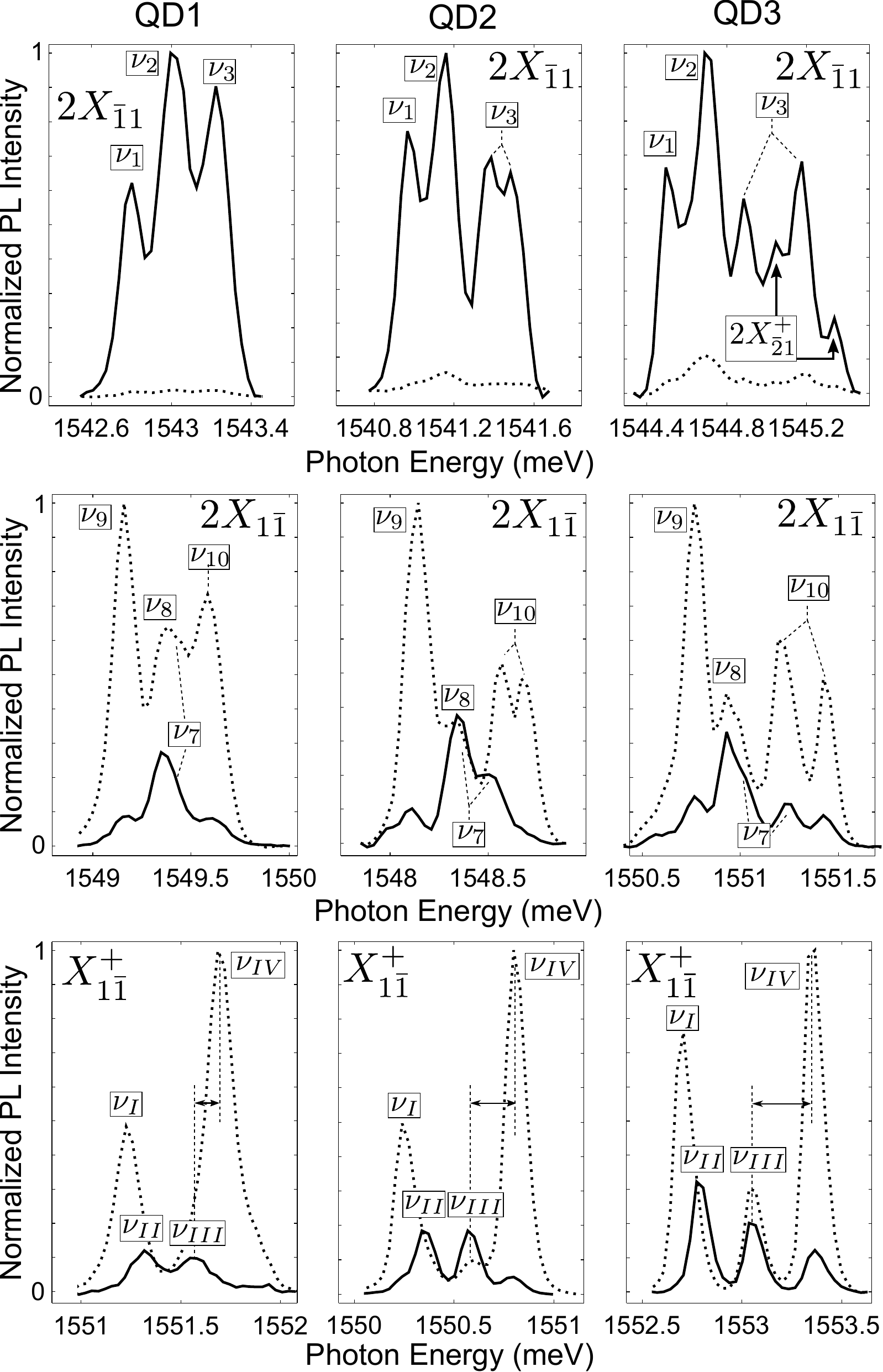}
\caption{Experimental $\mu$PL spectra of indicated exciton complexes acquired in side-view geometry organised in columns for three different QDs (QD1, QD2 and QD3). Solid (dotted) curves correspond to $y$-polarized ($z$-polarized) PL. Black thin vertical dotted lines aid the labeling of individual emission lines. QD1 to QD3 exhibit a successively stronger breaking from the ideal $C_{3v}$ symmetry.}
\label{sym_evol}
\end{figure}

The fine structure of about 15\% of the investigated QDs is well explained by the symmetry analysis of the previous section assuming a $C_{3v}$ point group. Other QDs in the same sample exhibit, however, more complex fine structure patterns. This is illustrated in Fig. ~\ref{sym_evol} for three different QDs (QD1-QD3) and the three transition patterns corresponding to $2X_{\bar{1}1}$, $2X_{1\bar{1}}$ and $X^+_{1\bar{1}}$. These transitions were previously shown to exhibit the effects of symmetry elevation towards $D_{3h}$. Here, the spectra of the dot labeled QD1 feature all the transitions derived in the previous section, while the spectra of QD2 and QD3 do exhibit additional emission lines and deviates from the strict polarization selection rules established for $C_{3v}$.

\begin{figure}[top] 
\includegraphics[width= 8.5 cm]{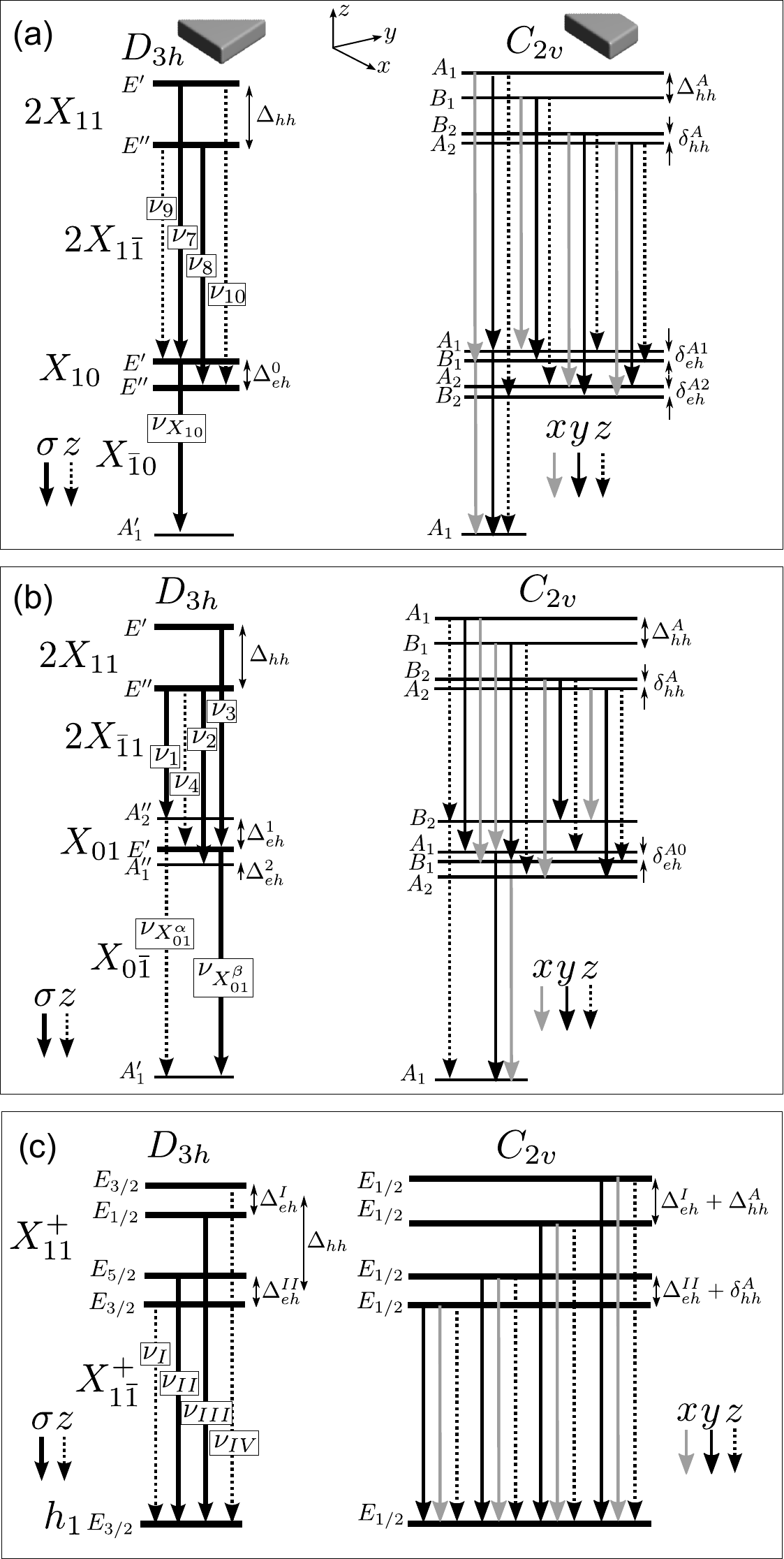}
\caption{ Group theory derived decay schemes of indicated complexes under $D_{3h}$ and $C_{2v}$. Doubly degenerate (non-degenerate) levels are shown with thick (thin) horizontal lines. Transitions with isotropic polarization in the $xy$-plane are represented by thick vertical lines, $x$-polarized ($y$-polarized) and $z$-polarized transitions are represented by grey (black) solid and black dotted vertical lines, respectively. }
\label{dec_2X11_C2v}
\end{figure}

In order to analyze the effects of symmetry breaking on the fine structure patterns, it is for simplicity assumed that breaking occurs from the elevated symmetry $D_{3h}$ towards the elevated symmetry $C_{2v}$ (from $C_s$). Here, one symmetry plane of $C_{2v}$ is perpendicular to the $xy$-plane, which is in contrast to the standard geometry of SK QDs where both symmetry planes of $C_{2v}$ are perpendicular to the $xy$-plane. In Fig. ~\ref{dec_2X11_C2v} we show the decay schemes of the three patterns under discussion, which are derived with an analysis assuming a $C_{2v}$ point group. For convenience, the corresponding schemes for $D_{3h}$ are repeated in the figure. The symmetry breaking towards $C_{2v}$ has three main effects on the spectra: i) Degenerate levels split into two energy levels. ii) The isotropic polarization associated with degenerate levels splits into $x$- and $y$-polarized components, i.e. the polarization \emph{isotropy} is lost. iii) New transitions become optically activated, in particular new $z$-polarized transitions appear for $X_{\bar{1}0}$, $2X_{\bar{1}1}$ and $2X_{1\bar{1}}$, while all the $X^+_{1\bar{1}}$ transitions are allowed for all the polarization directions. Note that any further symmetry breaking, e.g. towards $C_s$, would not cause any additional energy splitting, but more emission lines would become optically activated in other polarization directions.

Inspection of the top $2X_{\bar{1}1}$ spectra in Fig. ~\ref{sym_evol} suggests that the main difference between QD1 and QD2 (QD3) is a splitting of about $\sim$120 $\mu$eV ($\sim$280 $\mu$eV) for transition $\nu_3$. Any energy splitting of the spectral line $\nu_3$ in an asymmetric dot could originate either from a splitting of its initial state, i.e. the upper $E'$ state of $2X_{11}$, caused by h-h exchange interactions ($\Delta^A_{hh}$), or from a splitting of its final state, the $E'$ state of $X_{01}$, caused by e-h interactions ($\delta^{A0}_{eh}$) [see Fig. ~\ref{dec_2X11_C2v} (b)]. Note that the h-h exchange interaction energy $\Delta^A_{hh}$ would also cause splittings of transitions of $\nu_{10}$ and $\nu_7$ of $2X_{1\bar{1}}$, which originate from the same initial state. Splittings with comparable values for these two transitions are indeed observed for the pattern of $2X_{1\bar{1}}$ in the experimental spectra of QD2 and QD3 that are reported in Fig. ~\ref{sym_evol} (middle row). Since no splittings are resolved for the other transitions ($\nu_1$, $\nu_2$, $\nu_8$ and $\nu_9$) originating from the lower $E''$ state of $2X_{11}$, it must be concluded the h-h exchange splitting of that state under asymmetry ($\delta^A_{hh}$) is too small to be resolved for these QDs. It can also be concluded that any further splitting caused by e-h exchange interactions in the final single exciton states under asymmetry ($\delta^{A0}_{eh}$, $\delta^{A1}_{eh}$ and $\delta^{A2}_{eh}$) remains unresolved in the spectra of Fig. ~\ref{sym_evol}.

\begin{figure}[top] 
\includegraphics[width= 8.0 cm]{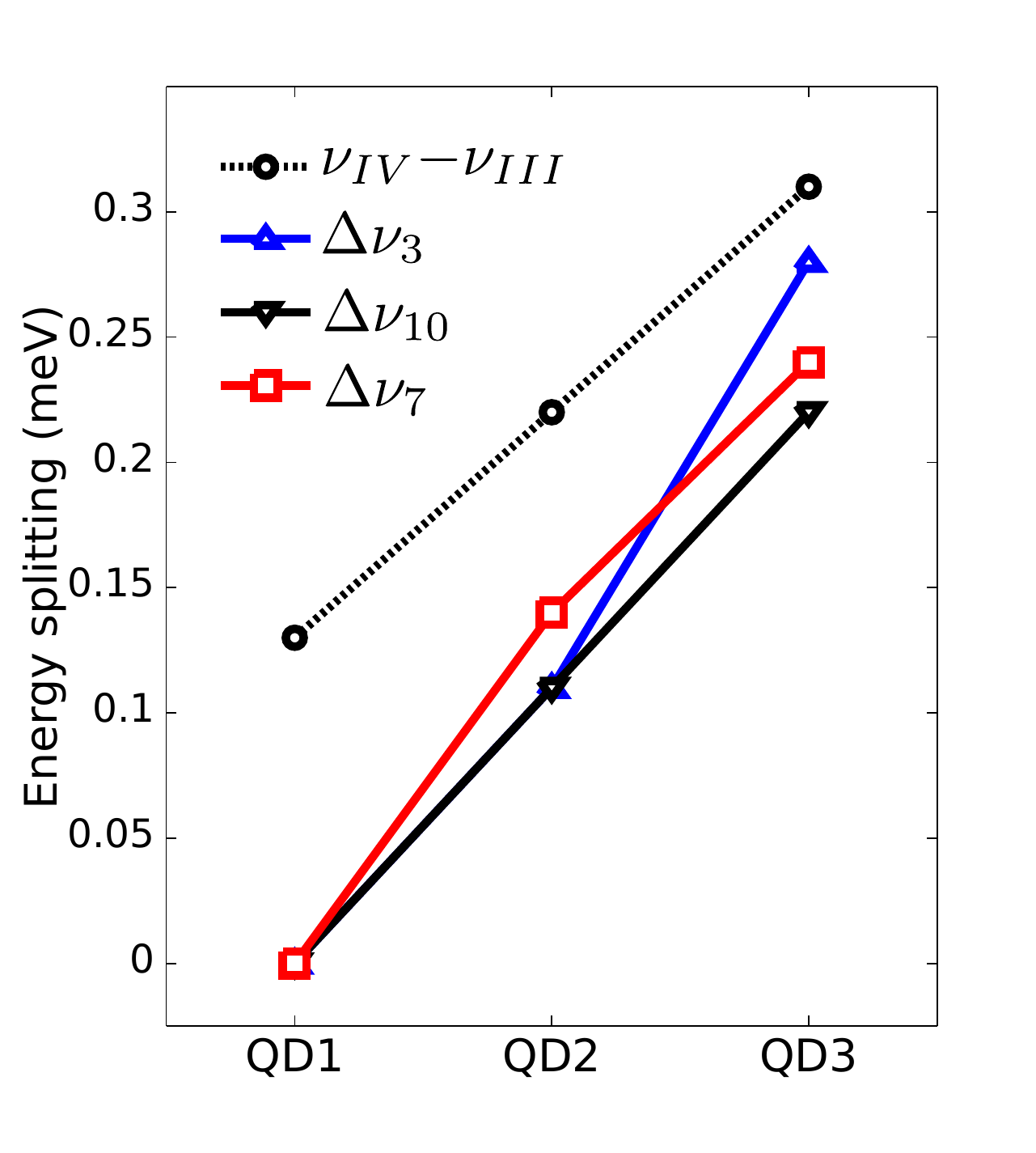}
\caption{(Color online) Measured energy splittings of $\nu_3$, $\nu_7$ and $\nu_{10}$, as well as the energy separation between $\nu_{IV}$ - $\nu_{IV}$ for the three QDs of Fig. ~\ref{sym_evol} exhibiting successively stronger breaking from the ideal $C_{3v}$ symmetry. For QD1, the splittings of $\nu_3$, $\nu_7$ and $\nu_{10}$ are unresolved, with the corresponding values set to zero.}
\label{asym_split}
\end{figure}

Based on the arguments given above, the interpretation of the spectra of QD2 and QD3 is that these dots exhibit successively stronger breaking from the ideal $C_{3v}$ symmetry (with $D_{3h}$ elevation). One dominating signature of symmetry breaking for biexcion $2X_{11}$ was identified as the h-h exchange induced splitting $\Delta^A_{hh}$ of its excited $E'$ state. h-h interactions of the same origin exist also for the trion $X^+_{11}$, but the e-h interactions of this complex already result in split excitonic states for a symmetric QD. However, for asymmetric QDs, the additional splitting $\Delta^A_{hh}$ due to h-h interactions will be superimposed onto existing splittings caused by e-h-interactions, modifying the energy spacing between the $\nu_{IV}$ and the $\nu_{III}$ transitions originating from the split hole levels. It can be seen in the $X^+_{1\bar{1}}$ spectra of Fig. ~\ref{sym_evol} (last row), going from QD1 to QD3, that the energy spacing successively increases for the two transitions identified as $\nu_{IV}$ and the $\nu_{III}$, with an amount comparable to the observed energy splittings of $2X_{11}$. This trend is more clearly represented in Fig. ~\ref{asym_split} (a) where all the values of the discussed energy splittings are plotted. 

It should also be noted that the set of hh-like transitions $2X_{\bar{1}1}$ gains stronger $z$-polarization with increased symmetry breaking [see Fig. ~\ref{sym_evol} (top row)]. This is consistent with opening of new $z$-polarized decay channels, but it also indicates that the relatively small lh-character associated with the hh-like level $h_1$ is responsible for an increased overlap with the electron wave function for asymmetric quantum dots. Similar relaxation of the polarization selection rules are also observed for the set of lh-like transitions $X^+_{1\bar{1}}$ [see Fig. ~\ref{sym_evol} (bottom row)]. For a symmetric dot, under $C_{3v}$ as well as $D_{3h}$, strict polarization rules were predicted with $\nu_I$ and $\nu_{IV}$ as $z$-polarized while $\nu_{II}$ and $\nu_{III}$ are $\sigma$-polarized transitions. For a QD with $C_{2v}$ symmetry (or lower), all transitions are active in all polarization directions. It can clearly be seen for QD3 that $\nu_{IV}$ has gained  in intensity in the $y$-polarized direction and that $\nu_{III}$ appears also with $z$-polarized light. Interestingly, the strict polarization rules of $C_{3v}$ still hold for $\nu_{I}$ and $\nu_{II}$, which are the transition originating from the lower $X^+_{11}$ levels that were found previously to be nearly unaffected by the symmetry breaking.

\begin{figure}[top] 
\includegraphics[width= 8.0 cm]{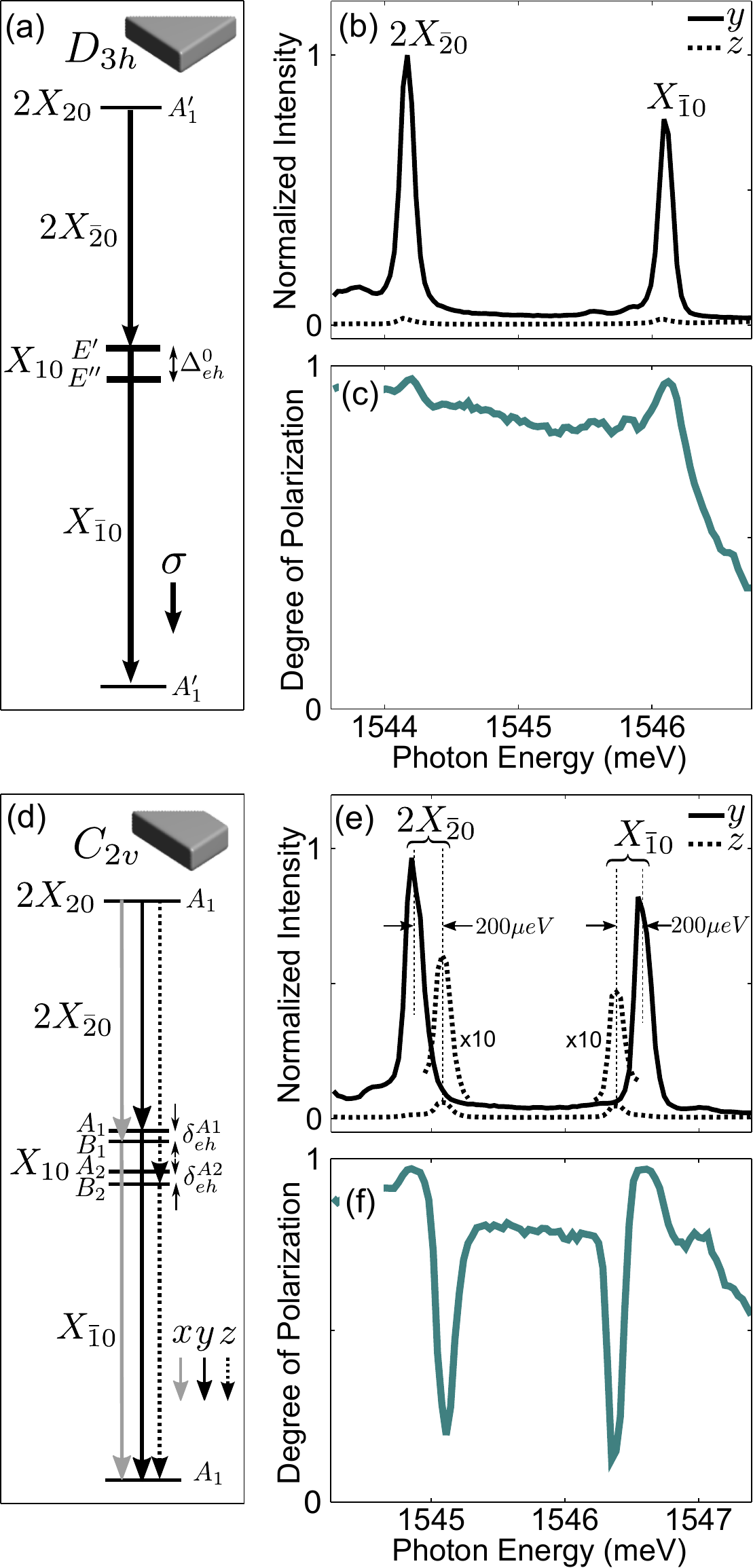}
\caption{(a) Group theory derived decay schemes of the conventional biexciton $2X_{20}$ under $D_{3h}$. Doubly degenerate (non-degenerate) levels are shown with thick (thin) horizontal lines. Transitions with isotropic polarization in the $xy$-plane are represented by thick vertical lines. (b) $\mu$PL spectra of the conventional exciton $X_{10}$ and biexciton $2X_{20}$ of a symmetric QD acquired in side-view geometry. The solid (dotted) black lines indicate PL intensity $I_y$ ($I_z$) linearly polarized along $y$ ($z$). (c) The degree of linear polarization $P = (I_y-I_z)/(I_y+I_z)$. (d) Analogous to (a) but for $C_{2v}$. $x$-polarized ($y$-polarized) and $z$-polarized transitions are represented by grey (black) solid and black dotted vertical lines, respectively. (e) and (f) Analogous to (b) and (c) but for an asymmetric QD.}
\label{sym_asym_dop}
\end{figure}

An additional consequence of symmetry breaking towards $C_{2v}$ is the activation of a $z$-polarized transition for the hh-like single exciton $X_{10}$ [see Fig. ~\ref{dec_2X11_C2v} (a)]. As this exciton is the final state in the decay of the biexciton $2X_{20}$, any corresponding $z$-polarized component should be observed for $2X_{\bar{2}0}$. 	Decay schemes of $2X_{20}$ for both $D_{3h}$ and  $C_{3v}$ are shown in Figs. ~\ref{sym_asym_dop} (a) and ~\ref{sym_asym_dop} (d), respectively. As already mentioned, the corresponding PL-spectrum of $2X_{\bar{2}0}$ and $X_{\bar{1}0}$ of the symmetric QD1 reported in Fig. ~\ref{dec_2X11_C2v} (b) do not reveal any $z$-polarized components. On the other hand, the spectra of the asymmetric QD3 shown Fig. ~\ref{dec_2X11_C2v} (e) evidences weak $z$-polarized components about 200 $\mu$eV below (above) the main $y$-polarized transition of $X_{\bar{1}0}$ ($X_{\bar{2}0}$). These $z$-polarized components yield significant dominating dips in the corresponding degree of polarization plotted in Fig. ~\ref{sym_asym_dop} (d). The energy spacing between $y$- and $z$-polarized components of $\sim$ 200 $\mu$eV is nearly the same as the previously estimated splitting $\Delta^0_{eh}$ = 172 $\mu$eV between the bright and the approximately dark state of $X_{10}$ for a symmetric QD. A small differences of the splittings are expected between symmetric and weakly asymmetric QDs due to the symmetry dependencies of the interactions energies, e.g. due to the symmetry induced splittings $\delta_{eh}^{A1}$ and $\delta_{eh}^{A2}$ [see Fig. ~\ref{dec_2X11_C2v} (d)]. These results demonstrate that $C_{2v}$ QDs do not exhibit two dark states, but merely a single one. Note that $z$-polarized components is normally not accessible in conventional top-view light collection, where the $x$- and $y$-polarized components are probed.

The side-view light collection implemented in this analysis of symmetry breaking may be inconvenient for the characterization of a large numbers of QDs, since only a one-dimensional array of dots can be accessed in this geometry. Instead the conventional top-view geometry may be more suitable for fast and efficient symmetry characterization, with access to two-dimensional arrays of QDs. Two of the strong signatures of symmetry breaking described so far are, however, clearly observed in the standard top view geometry, even without the need of polarization resolved spectroscopy in the $xy$-plane: (1) The detection of a splitting of the $\nu_3$ transition of $2X_{\bar{1}1}$, i.e. the observation of four instead of three peaks in the spectral pattern of $2X_{\bar{1}1}$, and (2) the detection of a third peak corresponding to $\nu_{IV}$ of $X^+_{1\bar{1}}$, i.e. the observation of more that two peaks in the spectral pattern of $X^+_{1\bar{1}}$. The first case should be considered the most robust, since it is insensitive to polarization cross-talk which may occur for certain sample structures for which a small amount of $z$-polarized light can be also detected from the top-view due to light scattering.

The signatures of symmetry breaking described in this section are much less demanding to resolve than the standard measurements of e-h exchange induced fine structure splitting $\delta_{eh}^{A1}$, which has typical values in the range of 0 - 50 $\mu$eV \cite{PhysRevB.80.165312}. The sensitivity of h-h exchange interaction to a weak symmetry breaking was clearly evidenced in the optical spectra of the biexciton $2X_{11}$, yielding splittings $\Delta_{hh}^A$ up to 250 $\mu$eV. This higher sensitivity suggests that a simple measurements of $\Delta_{hh}^A$ can be performed, instead of the direct measurement of $\delta_{eh}^{A1}$, in order to assess the symmetry and efficiently select of QDs suitable for certain experiments or applications  (e.g. for polarization entangled photon sources). 
\
\section{Discussion}
We will highlight some direct consequences of the analysis made in this work and discuss them in relation experimental and theoretical studies, mainly performed on conventional SK QDs In(Ga)As/GaAs QDs. Parts of the points discussed here have briefly been mentioned in our previous  publications, in which we did not present a complete analysis of the exciton complexes \cite{PhysRevB.81.161307, PhysRevLett.107.127403}.

The direct access to the energy separation between $h_2$ and $h_1$ was originally reported by Siebert et al. from the optical spectra of the excited positive trion in InAs/GaAs SK QDs \cite{PhysRevB.79.205321}. In that case, the relevant transitions were, however, determined by resonant excitation and PL-excitation spectroscopy. In our work, we perform a non-resonant $\mu$PL spectroscopy of $X^+_{11}$ and we could find that there was only a small reduction of the exciton binding energy of $\sim$0.6 meV for the excited lh-like exciton in comparison to its hh-like counterpart. The corresponding binding energy difference reported for InAs/GaAs SK QDs covers a wider range, however, from 0.5 to 2.7 meV \cite{PhysRevB.79.205321}. For these SK QDs the ground and excited hole levels have a dominant hh-character.  In both cases, the reduction of binding energy is understood as a reduction of the attractive e-h interaction, which takes place because the excited hole state has a smaller overlap with the electron state than the ground hole state.

It is widely accepted that the single exciton of a QD exhibit two bright states and two dark states \cite{PhysRevB.65.195315}. This description of the fine structure of an exciton underlines that the dark states are characterized by a total angular momentum $J$ of $\pm$2 arising from the electron ($\pm$1/2) and hole spins ($\pm$3/2) in a zinc-blend semiconductor heterostructure \cite{PSSA:PSSA487}. Our group theoretical analysis of the optical selection rules reveals, however, that a QD with $C_{2v}$ symmetry exhibit a single dark state [see Fig. ~\ref{sym_asym_dop} (d)], while the three other exciton states are optically active with $x$-, $y$- and $z$-polarization, respectively \cite{PhysRevB.81.161307, PhysRevLett.107.127403}. This prediction of a $z$-polarized component is indeed confirmed by the experimental results displayed in Figs. ~\ref{sym_asym_dop} (e) and (f), where this $z$-polarized spectral line is positioned 200 $\mu$eV from the single exciton transition $X_{\bar{1}0}$. On the other hand, the existence of a $z$-polarized component in the emission spectrum of an exciton was experimentally demonstrated for asymmetric self-assembled CdTe/ZnTe QDs \cite{PhysRevB.86.241305} and theoretically confirmed using a tight-binding model \cite{PhysRevB.87.115310}.

On the basis of a simple single band model it was predicted that the fine structure of excited trions has a universal structure for both positively and negatively charged excitons; the main structure originates from the singlet-triplet states caused by exchange interaction between pairs of holes or electrons \cite{PSSA.3.592}. However, experiments as well as atomistic theory on double charged exciton in $C_{2v}$ InAs/GaAs QD demonstrated a clear distinction between electron-electron and hole-hole exchange interaction. While two electrons form singlet-triplet states, two holes instead form one doublet and two singlets \cite{PhysRevLett.98.036808}. The explanation for this difference is drawn on the spinorial nature of the single-particle states: spin 1/2 for the electrons and spin 3/2 for the holes.
For heavy-holes this would lead to a twofold degenerate state with $J$ = 3 and two singlets with $J$ = 0 and $J$ = 2 \cite{PhysRevLett.98.036808}. On the basis of the group theory approach used in our work we could predict that for QDs with $C_{3v}$ symmetry, the two holes only belong to two doubly-degenerate states. This prediction is the origin of the two $E$ states of $2X_{11}$. Furthermore, it was found that under $C_{2v}$ any pair of holes (or electrons) under exchange interaction corresponds to non-degenerate states [see top part of Fig. ~\ref{dec_2X11_C2v}]. These results hold independently of the heavy or light hole character of the single-particle states.

The maximal h-h exchange interaction energies $\Delta_{hh}$ determined for the pyramidal QDs in this work are typically below 300 $\mu$eV. This small value should be compared to the e-e exchange interaction $\Delta_{ee}$ of more than 7 meV that was experimentally determined from the emission lines of a double negatively charged exciton for similar QDs with excited electron states \cite{PhysRevLett.84.5648}. The tiny value of $\Delta_{hh}$ reported here can appear surprising on a first glance, especially when considering that the holes are heavier and thus more localized than the electron, which would lead to larger Coulomb interaction energies for holes than for electrons. However, exchange interactions only occurs between identical particles. For pure hh- and lh-states, no long-range h-h exchange splitting exist at all due to the spinorial orthogonality between the two states. The fact that holes in $h_1$ are to a large extent hh ($\sim$90 \%) while holes in $h_2$ are only $\sim$10 \% hh indeed leads to very small values of  $\Delta_{hh}$ in the investigated QDs. Thus, much larger values would be expected if the character of the holes in $h_1$ and $h_2$ were similar, or if both were to be strongly hh-lh mixed. Corresponding values of $\Delta_{hh}$ for conventional SK QDs, where both hole levels involved have mainly a hh-character, have been determined to be in the range of 2 to 5 meV \cite{PhysRevB.79.125316, PhysRevLett.98.036808}. In other QDs with hole states of similar hole character or with higher degree of hh-lh mixing, the h-h exchange induced splitting $\Delta_{hh}^A$ upon symmetry breaking can be much more significant than in the present work, with values in the meV range \cite{PhysRevB.88.045321}.

\section{Conclusions}
A sequence of experimental methods has been implemented in order to reach a complete understanding of the various intrinsic spectral features of exciton complexes in pyramidal QDs. Our approach enables a solid experimental identification of all relevant exciton complexes, including complexes with weak emission lines and overlapping spectral features  without relying on any quantitative estimation of their fine structure splittings.  A general group theoretical analysis of the excitonic fine structure was presented for all the observed exciton complexes, which enabled the the identification and determination of the e-h and h-h exchange interaction energies contributing to the experimentally observed splittings. Furthermore, it was shown that the observed dot-to-dot variation of the fine structure splittings could be consistently explained as effects of various degrees of breaking from the ideal $C_{3v}$ symmetry. This work emphasizes the importance of the actual QD symmetry as well as approximate elevated symmetries in order to derive the optical selection rules of the exciton complexes and to reach a comprehensive description of their emission patterns.

\bibliography{spectralSignatures_KFKarlsson}

\end{document}